\documentclass[aps,pre,twocolumn,groupedaddress,10pt]{revtex4}

\usepackage[english]{babel}
\usepackage[T1]{fontenc}
\usepackage{graphicx}
\usepackage{psfrag}
\usepackage{color} 
\usepackage{bbm}
\usepackage{amsmath}
\usepackage{hyperref}
\hypersetup{pdfstartview={FitH}}

\newcommand{\bq}{\begin{equation}}
\newcommand{\eq}{\end{equation}}
\newcommand{\ba}{\begin{eqnarray}}
\newcommand{\ea}{\end{eqnarray}}

\begin{document} 

\title{Resonant noise amplification in a predator-prey model with quasi-discrete generations}
\author{M. Giannakou$^{1,3}$, B. Waclaw$^{1,2}$}

\affiliation{$^1$School of Physics and Astronomy, University of Edinburgh, James Clerk Maxwell Building, Peter Guthrie Tait Road, Edinburgh, EH9 3FD, United Kingdom\\
$^2$Dioscuri Centre for Physics and Chemistry of Bacteria, Institute of Physical Chemistry PAS, Kasprzaka 44/52, 01-224 Warsaw, Poland\\
$^3$Institut für Physik, Johannes Gutenberg-Universität Mainz, Staudingerweg 9, 55128 Mainz, Germany}

\begin{abstract}
	Predator-prey models have been shown to exhibit resonance-like behaviour, in which random fluctuations in the number of organisms (demographic noise) are amplified when their frequency is close to the natural oscillatory frequency of the system. This behaviour has been traditionally studied in models with exponentially distributed replication and death times. Here we consider a biologically more realistic model, in which organisms replicate quasi-synchronously such that the distribution of replication times has a narrow maximum at some $T>0$ corresponding to the mean doubling time. We show that when the frequency of replication $f=1/T$ is tuned to the natural oscillatory frequency of the predator-prey model, the system exhibits oscillations that are much stronger than in the model with Poissonian (non-synchronous) replication and death. The effect can be explained by resonant amplification of coloured noise generated by quasi-synchronous replication events. 
	To show this, we consider a single-species model with quasi-synchronous replication. We calculate the spectrum and the amplitude of demographic noise in this model, and use these results to obtain these quantities for the two-species model.
\end{abstract}

\maketitle

\section{Introduction}	
When a non-linear dynamical system capable of exhibiting damped oscillations  is coupled to a source of random noise, it often begins to generate periodic oscillations (a quasi-cycle) \cite{boland_how_2008}. This resonance-like behaviour is caused by the amplification of noise frequencies that are in tune with the natural oscillatory frequency of the system. Importantly, the noise does not have to be external but it can be intrinsic to the system and arise from its microscopic stochastic dynamics. 

An important example is resonant amplification of demographic noise which has been found in stochastic models of biological populations \cite{mckane_predator-prey_2005,pineda_krch_tale_2007, mckane_amplified_2007,rozhnova_stochastic_2010,huang_stochastic_2015,dobramysl_stochastic_2018,karig_stochastic_2018}. However, all these models assume that reproduction is a Markov process: birth and death occur with certain (possibly state-dependent) rates. At any moment, the distribution of replication times is therefore exponential, with the maximum at $t=0$. However, biological organisms do not replicate in this way: all known organisms require a certain minimum time to develop reproductive capability. Moreover, many organisms reproduce in quasi-discrete generations such that the time between consecutive replication events has a narrow distribution that peaks around some characteristic time $T$ called the generation time, or doubling time. For example, for the bacterium {\it E. coli}, $T$ ranges between 20 min and a few hours and the coefficient of variation of the doubling time is $0.1-0.3$, depending on growth conditions \cite{wallden_synchronization_2016,kennard_individuality_2016}. This leads to significant correlations between reproduction times of related individuals. Modelling this process for a single species has a long history \cite{powell_growth_1956,lebowitz_theory_1974,webb_model_1986,thomas_making_2017,lin_effects_2017,jafarpour_cell_2019,jedrak_generalization_2022}.

The quasi-synchronous nature of replication suggest an interesting possibility: if a predator-prey system has a tendency to oscillate at a frequency similar to the inverse of the doubling time, synchronisation of the two oscillations may lead to a substantial enhancement of resonant amplification of demographic noise.

In this work, we investigate this scenario in a simple predator-prey model originally proposed in Ref. \cite{mckane_predator-prey_2005}. The model assumes two biological species interacting in a way that leads to damped predator-prey cycles in the limit of infinitely large populations. In the original model, demographic noise due to stochastic replication of organisms led to persistent oscillations of small but non-zero amplitude and a Lorenz-like power spectrum. Here we show that when replication is no longer Poissonian but occurs in quasi-discrete generations, these oscillations increase dramatically in amplitude and can be as high as 50\% of the total population size even when the number of organisms is very large (millions or more).

\section{Model}
Our model an extension of the Newman-McKane model \cite{mckane_predator-prey_2005}. We consider a well-mixed population of two types of organisms A, B. We shall call these organisms "cells" as if they were single-celled microorganisms, although the model is agnostic to the exact nature of these organisms. Let $n_A,n_B$ be the number of cells of each type.
Each cell has an internal state variable $\tau$ assigned at birth from a certain distribution $R(\tau)$, the same for both species. We shall call this variable a ``timer''. The timer counts down from the assigned time interval; when it reaches zero, the cell produces an offspring and both cells are assigned new, randomly selected values of $\tau$ from $R(\tau)$. We shall assume that the distribution $R(\tau)$ is concentrated around its mean value $\left<\tau\right>\equiv T>0$. Cells also die with per-capita rates $d_A=p_2 -  p_1 n_B/K$ for type A, and $d_B=p_4 n_A/K  - p_3(1-n_B/K)$ for type B. Here $K$ plays a role similar to the carrying capacity in population dynamics models and sets the scale for the number of cells in the system: $n_A,n_B\sim K$ on average. We have used the same symbols for the parameters $p_1,p_2,p_3,p_4$ as in Ref. \cite{mckane_predator-prey_2005}, however their microscopic interpretation is slightly different. We will come back to this when we discuss the steady-state solution of the model.

The dynamics of the model can be schematically represented as a set of chemical-like equations:
\ba
	A \to 2A \label{eq:chem1} \\
	B \to 2B \\
	A \to 0 \\
	B \to 0	\label{eq:chem4}
\ea
However, one must be careful with how these equations are interpreted in our model. Only the last two reactions have been used in the same way as in Ref. \cite{mckane_predator-prey_2005} to represent inhomogeneous Poisson processes occurring with state-dependent rates $d_A(n_A,n_B),d_B(n_A,n_B)$. The first two reactions do not describe Poisson processes because the probability of replication depends on the internal state $\tau$ of each cell. Thus the model is non-Markovian.

Let us briefly discuss some possible choices for the distribution $R(\tau)$ of replication times.
The case $R(\tau)=\delta(\tau-T)$ corresponds to all cells reproducing in perfect synchrony; the generation time is $T$. The exponential distribution $R(\tau)=(1/T)\exp(-\tau/T)$ represents the Poisson case: cells reproduce with rate $1/T$ per capita and the mean time to replication is $T$. In this case the model is Markovian and its behaviour is expected to be the same as the original model from Ref. \cite{mckane_predator-prey_2005}. Finally, $R(\tau)$ can be concentrated around $\tau=T$ but have a non-zero width. This represents quasi-synchronous replication: all descendants of a given cell initially replicate in quasi-discrete generations, with progressive loss of synchronization over time.

In this manuscript, we shall compare the behaviour of the model for two distributions $R(\tau)$: (i) exponential (the Poisson model) with mean time to division $T$, (ii) uniform on $(T(1-w),T(1+w))$ where $w\ll 1$ controls the degree of correlation of replication times; synchrony is lost after $\sim 1/w$ generations.

\subsection{Biological interpretation}
We shall now provide a biological interpretation of the model. However, our analysis does not rely on this interpretation and the model is deliberately oversimplified and not intended to replicate any specific experiment. Our model could describe a population of micro-organisms of one species and two slightly different ecotypes. Both types replicate with the same time-independent rate. Type A's basal death rate $p_2$ decreases in the presence of B proportionally to the concentration of B times $p_1$. This could be due to type B producing an essential chemical compound necessary for A to survive. This interpretation in consistent with $p_2\gg b$ that we will generally assume later. Therefore, a sufficient density of B is required for A to thrive. Type B, on the other hand, is killed by A (e.g., A releases a toxin that kills B) with rate equal to the abundance of A times $p_4$. The additional term $-p_3(1-x_B)$ accounts for the increase in the death rate due to crowding. A similar term could be added to the equation for type A to make the model more symmetric but it would not qualitatively affect the dynamics of the model. All interactions described here have been demonstrated in microbial populations \cite{kerr_local_2002, smith_evolution_2020, pfeiffer_evolution_2004}.

\section{Analysis of the model}
We first study the behaviour of the Poisson model in the infinite-population size limit ($K\to\infty$) by neglecting fluctuations in the number of cells. We define $x_A=n_A/K, x_B=n_B/K$ as the new state variables. The dynamics of the model can be approximated by two differential equations:
\ba
	\frac{dx_A}{dt} &=& x_A \left(b - \max[ p_2 - p_1 x_B, 0] \right), \label{eq:dyn1}\\
	\frac{dx_B}{dt} &=& x_B \left(b - \max[p_4 x_A - p_3(1-x_B),0] \right), \label{eq:dyn2} 
\ea 	
where $b=\ln(2)/T$. In the above, we have used average rates of all processes represented by reactions (\ref{eq:chem1}-\ref{eq:chem4}), and assumed that all higher moments factorize into products of $x_A,x_B$. The $\max[\dots]$ function ensures that the death terms contribute only if the corresponding rates are positive. 

\begin{figure}
	\includegraphics[width=\columnwidth]{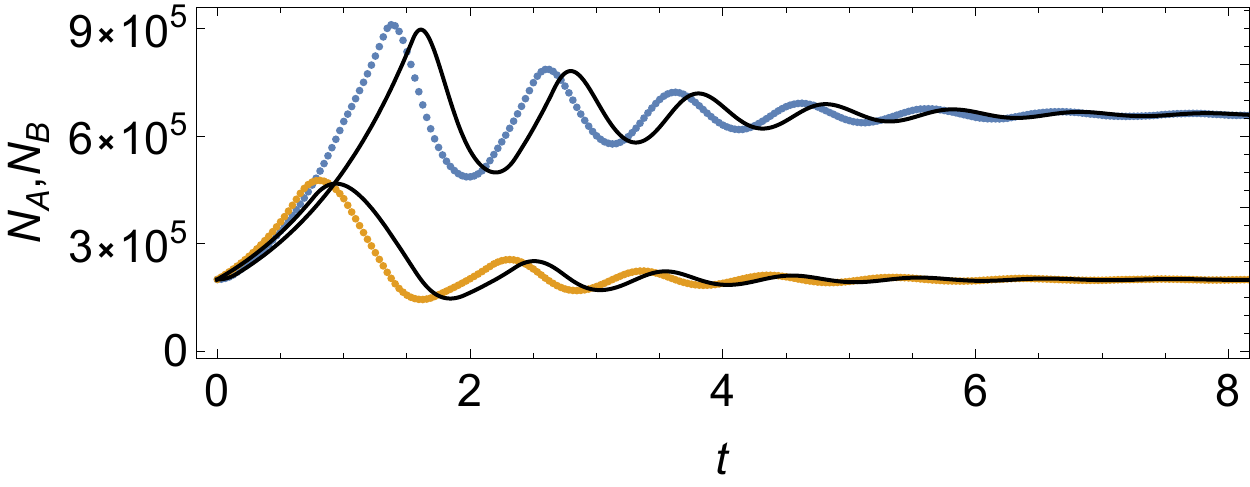}
	\includegraphics[width=\columnwidth]{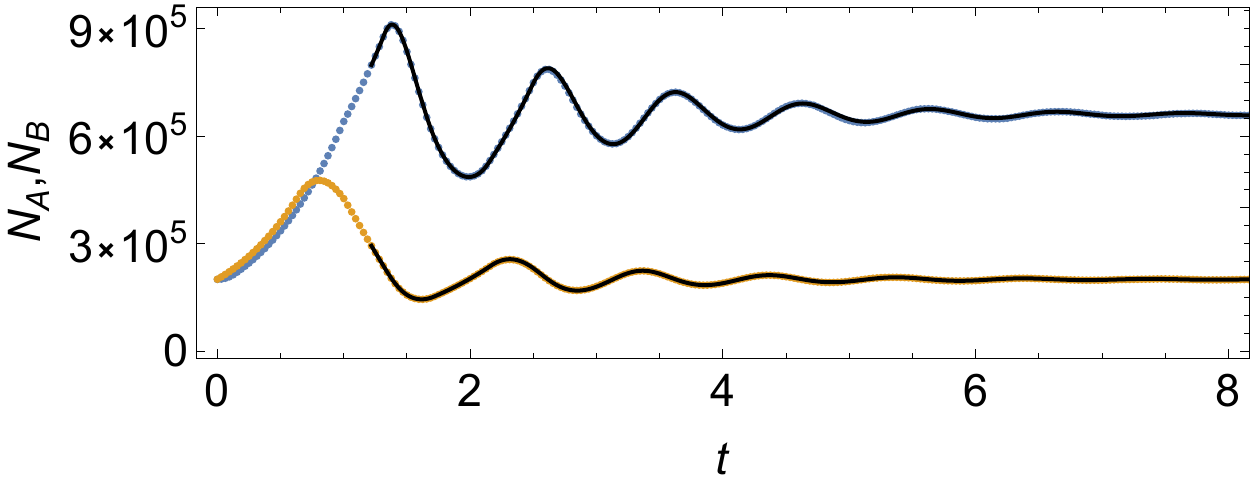}
	\caption{\label{fig:1}Damped oscillations in the stochastic asynchronous model (points: blue for $N_A$, yellow for $N_B$) and its deterministic counterpart (black lines), for  $\{p_1,p_2,p_3,p_4,b\}=\{30,7,7,10,1\}$. Top: the deterministic solution differs from the stochastic simulation for the uniform initial distribution of the timer variable. Both models assume the same initial condition $x_A(0)=0.2,x_B(0)=0.2$. Bottom: the agreement is very good when we compare the models after the stochastic model reached a quasi-steady state timer distribution. The deterministic model uses the values of $N_A,N_B=\{ 797423, 293636\}$ from the stochastic simulation for $t_0=1.2207$ as its initial condition. }
\end{figure}

The non-zero steady-state solution of the model, with both species present, reads:
\ba
  x_A^* &=& \frac{(p_1-p_2)p_3+b(p_1+p_3)}{p_1p_4} , \label{eq:ss1} \\
  x_B^* &=& \frac{p_2-b}{p_1}  . \label{eq:ss2}
\ea
To investigate the stability of this solution, we Taylor-expand  equations (\ref{eq:dyn1}-\ref{eq:dyn2}) around the steady-state solution. This leads to the Jacobian matrix with the following eigenvalues:
\begin{widetext}
\bq
	\lambda_\pm=\frac{b p_3 - p_2 p_3 \pm \sqrt{(b-p_2) \left(b (2 p_1 +p_3)^2 - p_3 \left(-4 p_1^2 + 4 p_1 p_2 + p_2 p_3\right)\right)}}{2 p_1}. \label{eq:lambdas}
\eq
\end{widetext}
In the regime that we are interested here ($x_A^*>0, x_B^*>0$), the eigenvalues have a negative real part, meaning that the steady-state solution is stable to a small perturbation. However, their imaginary part is generally non-zero, and hence the system will exhibit damped oscillations while relaxing towards the steady state. This is illustrated in Figure \ref{fig:1}, which shows plots of the deterministic solution for $\{p_1,p_2,p_3,p_4,b\}=\{30,7,7,10,1\}$, for a short time (a few oscillations), starting from $x_A(0)=0.2,x_B(0)=0.2$. On the same plot we show the results of a numerical simulation of the stochastic Poisson model with $K=10^6$. Note the small discrepancy between both models (Fig. \ref{fig:1}, top). This is due to the timer of all cells being initialized with uniformly distributed random numbers in the stochastic simulation. This initial distribution is quite different than the quasi-steady state distribution obtained after a few cycles, which is implicitly assumed when deriving Eqs. (\ref{eq:dyn1}, \ref{eq:dyn2}) from the microscopic rules (\ref{eq:chem1}-\ref{eq:chem4}). When we solve the deterministic model starting from a later point, using $N_A,N_B$ from the stochastic model as the initial condition, the discrepancy vanishes (Fig. \ref{fig:1}, bottom).

\subsection{Parameter selection}
The deterministic model has five parameters: $p_1,p_2,p_3,p_4,b$. We can put $b=1$; this fixes the time scale. The remaining four parameters determine the frequency of small-amplitude oscillations, the damping coefficient, and steady-state occupations. Before we move on, we shall discuss how we select these parameters so that the model exhibits under-damped oscillations; this is required for the resonant amplification of noise.
	
We have used Eqs. (\ref{eq:lambdas}) together with Eqs. (\ref{eq:ss1}-\ref{eq:ss2}) to find a region in the parameter space $\{p_1,p_2,p_3,p_4\}$ and $b=1$ of the deterministic model that corresponds to damped oscillations of frequency $f_0={\rm Im}(\lambda)/(2\pi)\in(0.98,1.02)$, damping coefficient $|{\rm Re}(\lambda)|<0.5$, and steady-state abundances $0.5\pm 0.1$. We did this via Monte-Carlo sampling of the parameter space. We then used the selected values as starting points for a root finding algorithm to find ${p_1,p_2,p_3,p_4}$ such that the frequency would be exactly $f_0=1$, steady-state occupations $x_A=x_B=0.5$, and the damping coefficient assumed one of three values: $0.5$ (fast damping), $0.2$ (slow damping) and $0.1$ (minimal damping).

This procedure has produced three sets of parameters: $S_{0.5}=\{39.73, 20.86, 2., 4.\}$, $S_{0.2}=\{56.45, 29.23, 0.8, 2.8\}$, and $S_{0.1}=\{65.81, 33.91, 0.4, 2.4\}$.
We shall use these parameters in the full, stochastic model.
	
\subsection{Numerical results}
\begin{figure}
	\includegraphics[width=\columnwidth]{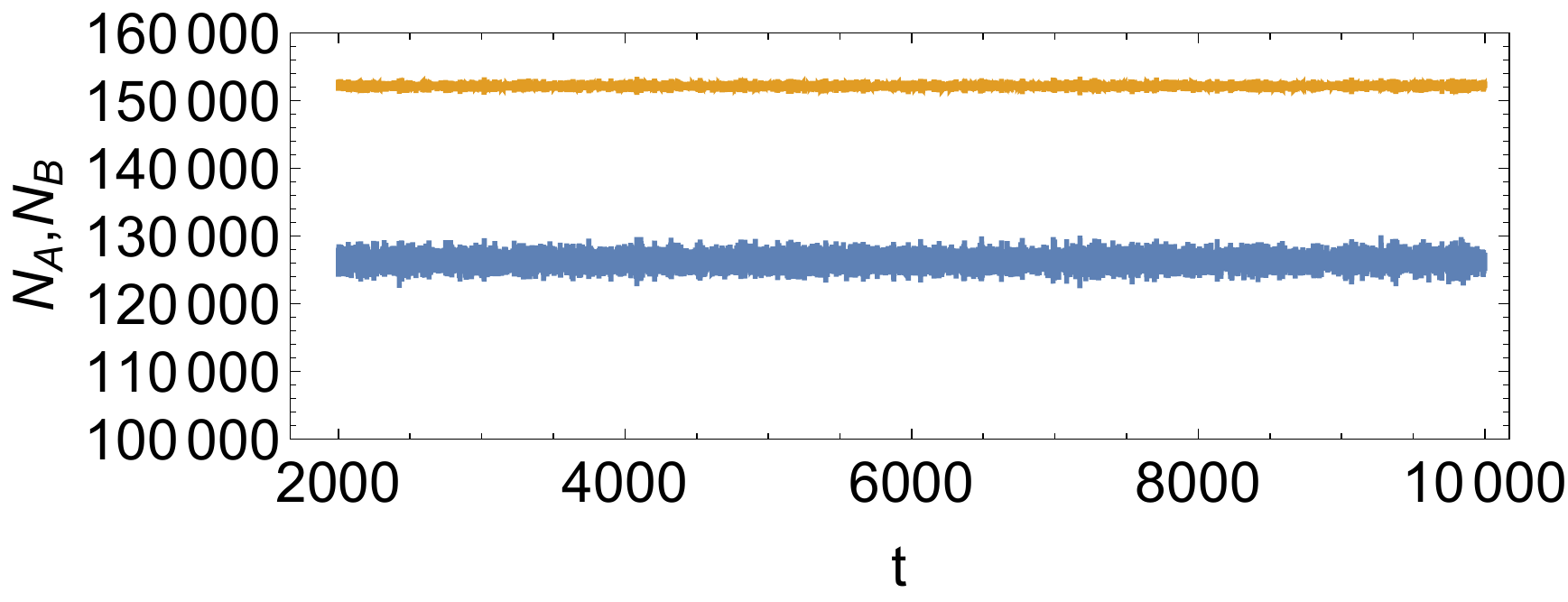}
	\includegraphics[width=\columnwidth]{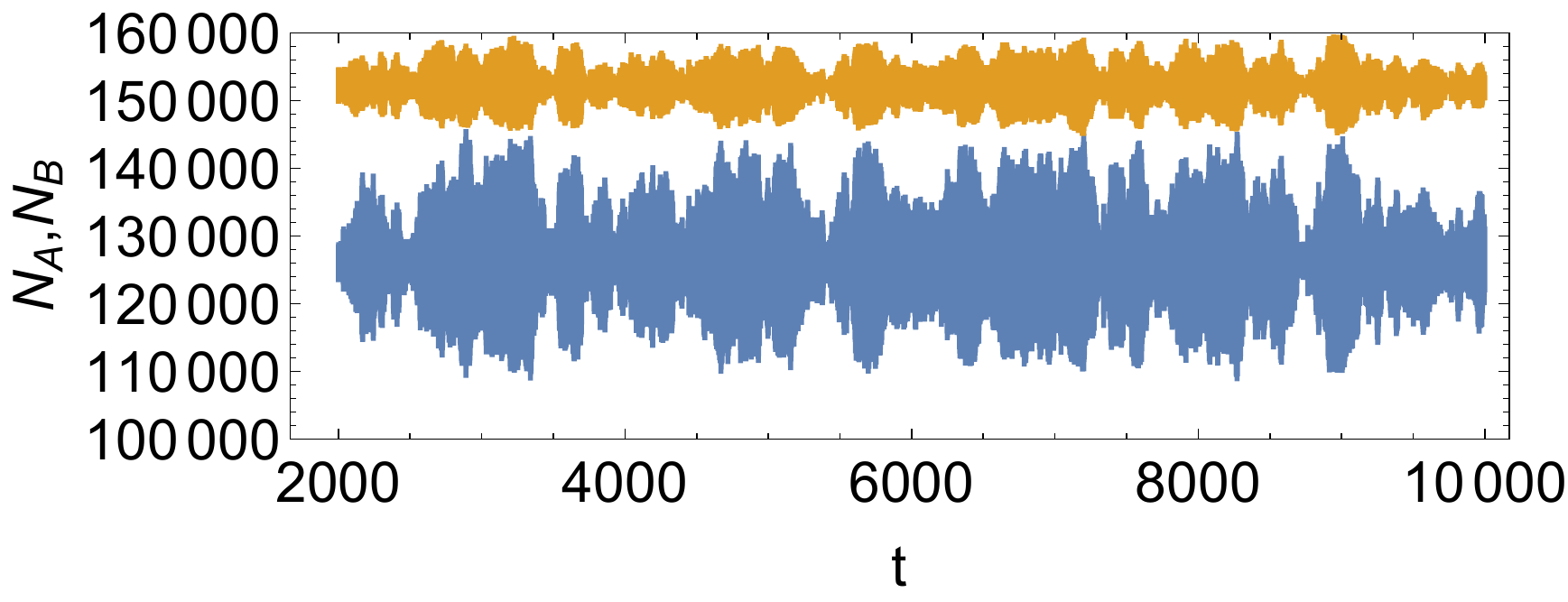}
	\caption{\label{fig:2}Example time series $N_A(t),N_B(t)$ in the Poisson (top) and quasi-synchronous (bottom) models, for $K=300000, w=0.02$. Individual oscillations cannot be seen due to the length of the time window shown here; the window contains a few thousand oscillations as those from Fig. \ref{fig:1}.}
\end{figure}

We have simulated the stochastic model using a simple tau-leaping algorithm with fixed-size time step $dt=1/512$ \cite{gillespie_approximate_2001}. Figure \ref{fig:2} shows examples of time series obtained for the Poisson version of the model, and for the quasi-synchronous model with a narrow distribution of doubling times. The parameters are $S_{0.5}$, $K=300000, w=0.02$, and the initial condition is $n_A(0)=0.2K, n_B(0)=0.2K$. In both cases the natural oscillatory frequency of the model is $f_0=1$ and the average doubling time is also $T=1$.

\begin{figure}
	\includegraphics[width=0.48\columnwidth]{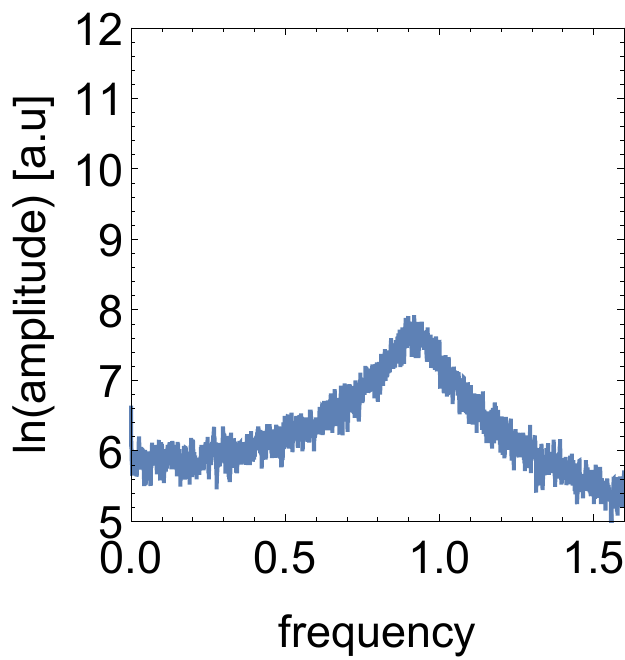}
	\includegraphics[width=0.48\columnwidth]{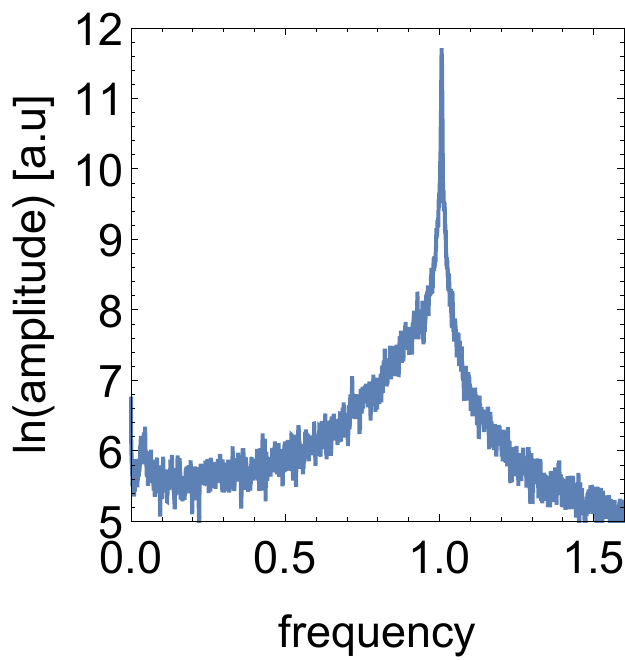}
	\caption{\label{fig:3}Fourier spectrum of $N_A(t)$ in the Poisson (left) and non-Poisson (right) models, for $K=300000,w=0.02$, the remaining parameters = $S_{0.5}$. Moving average with a 20-point long window has been applied to smooth out the spectra.}
\end{figure}

\begin{figure}[t!]
	\includegraphics[width=\columnwidth]{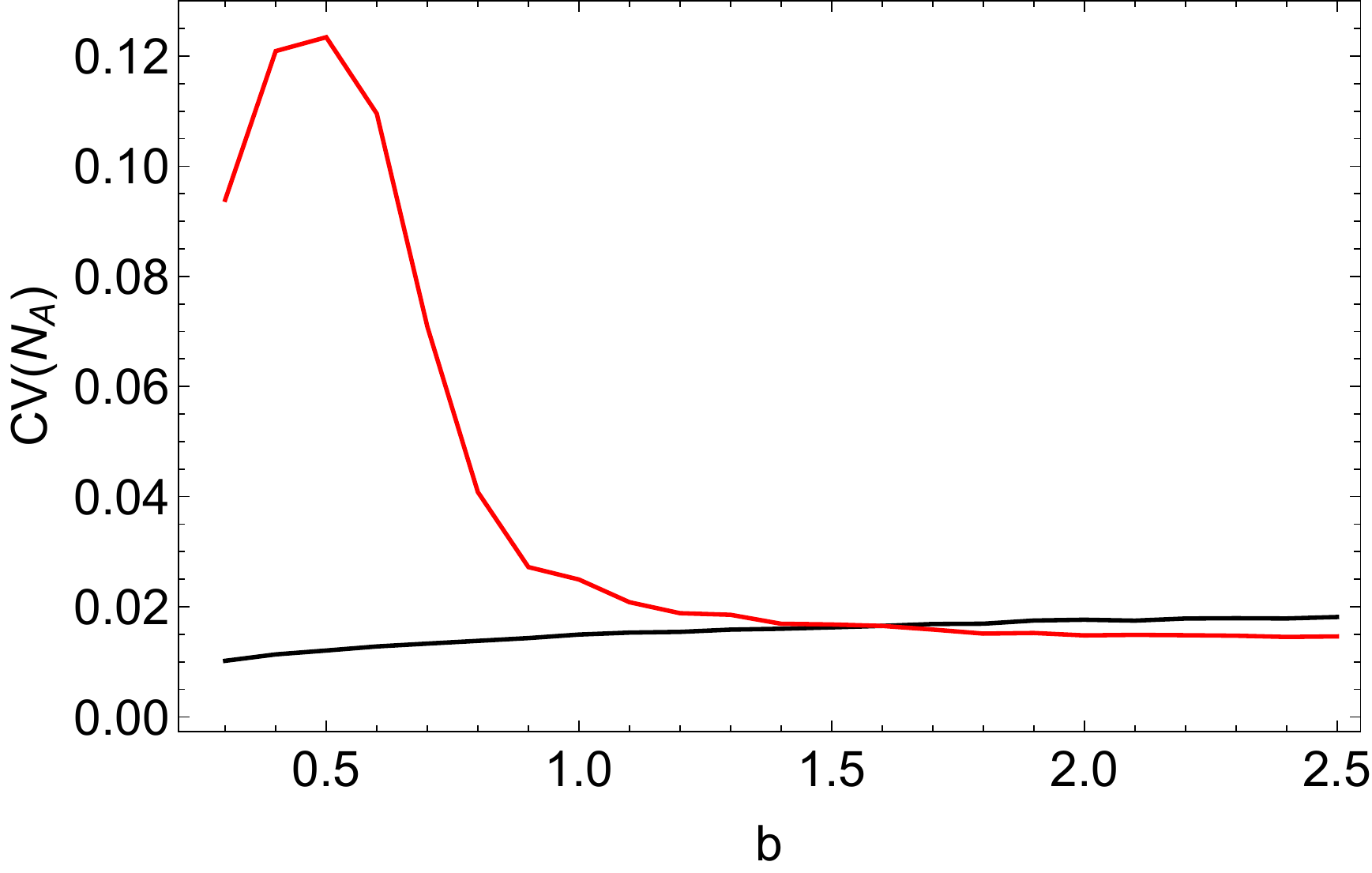}
	\includegraphics[width=\columnwidth]{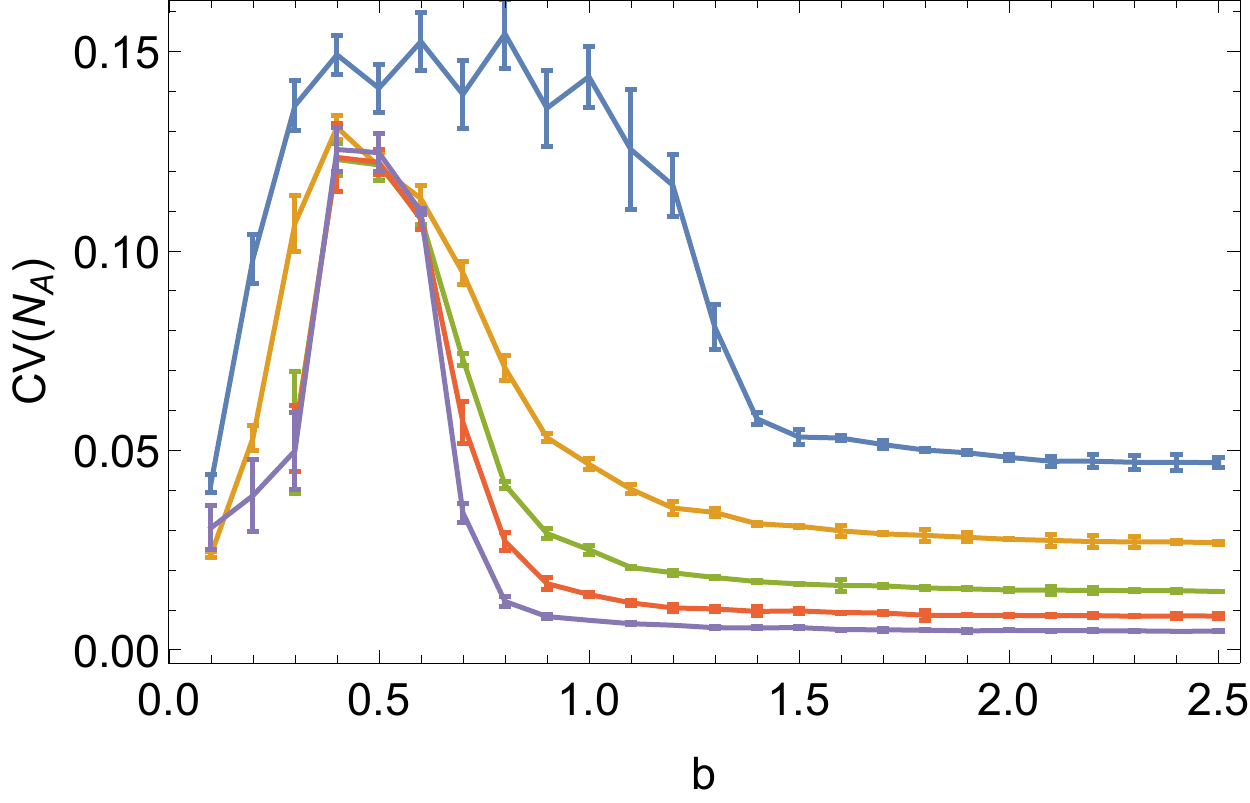}
	\includegraphics[width=\columnwidth]{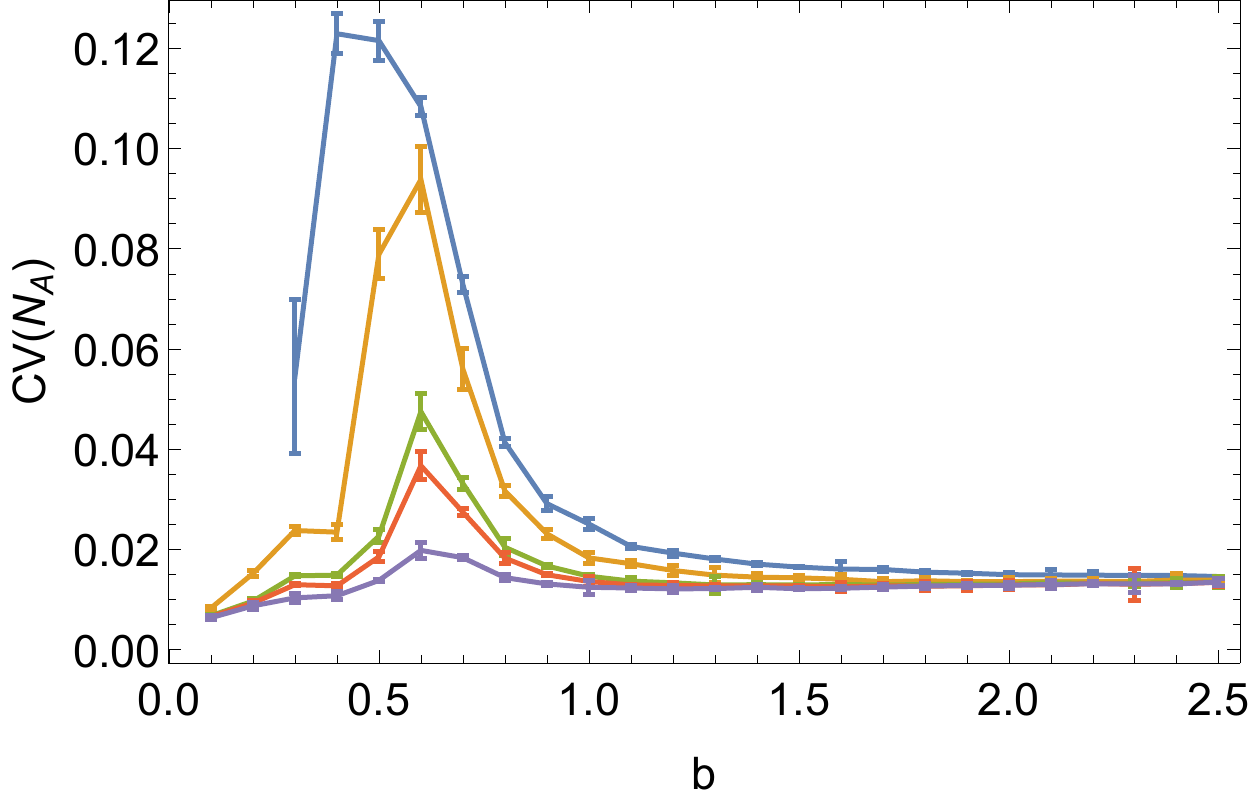}	
	\caption{\label{fig:4}Top: Coefficient of variation (CV) of $N_A(t)$ for the Poisson (black) and quasi-synchronous (red) models, as a function of $b$. CV is a convenient measure of the amplitude of oscillations. A resonance peak can be seen at $b\approx 0.5$. Middle: CV of $N_A(t)$ for the quasi-synchronous model with $w=0.02$ and different $K=10^4,3\times 10^4,10^5, 3\times 10^5, 10^6$ (blue, yellow, green, red, violet). Bottom: CV versus $b$ for different widths $w=0.02,0.04,0.08,0.1,0.2$ of the doubling time distribution (colours from blue to violet) and $K=10^5$. In all cases, parameters = $S_{0.5}$.}
\end{figure}

The quasi-synchronous  model exhibits much larger oscillations than the Poisson model. Accordingly, the Fourier spectrum of the quasi-synchronous  model has a much more pronounced peak at frequency $f_0=1$ (Fig. \ref{fig:3}).
The observed increase in the amplitude of oscillations occurs only when the doubling frequency is close to the natural oscillation frequency. Figure  \ref{fig:4}, top, shows that when $T\neq 1/f_0$ the amplitude is significantly reduced; this resonance-like behaviour is not present in the Poisson model (black line in Fig. \ref{fig:4}, top). Interestingly, the maximum amplitude is observed at a slightly lower $b=0.5$ than expected ($b=\ln 2=0.69$ which corresponds to $T=1$). The resonance peak is also quite broad. This is a non-linear effect; for large amplitudes as observed here, the resonant frequency is slightly lower than $f_0=1$. 

\begin{figure}
	\includegraphics[width=\columnwidth]{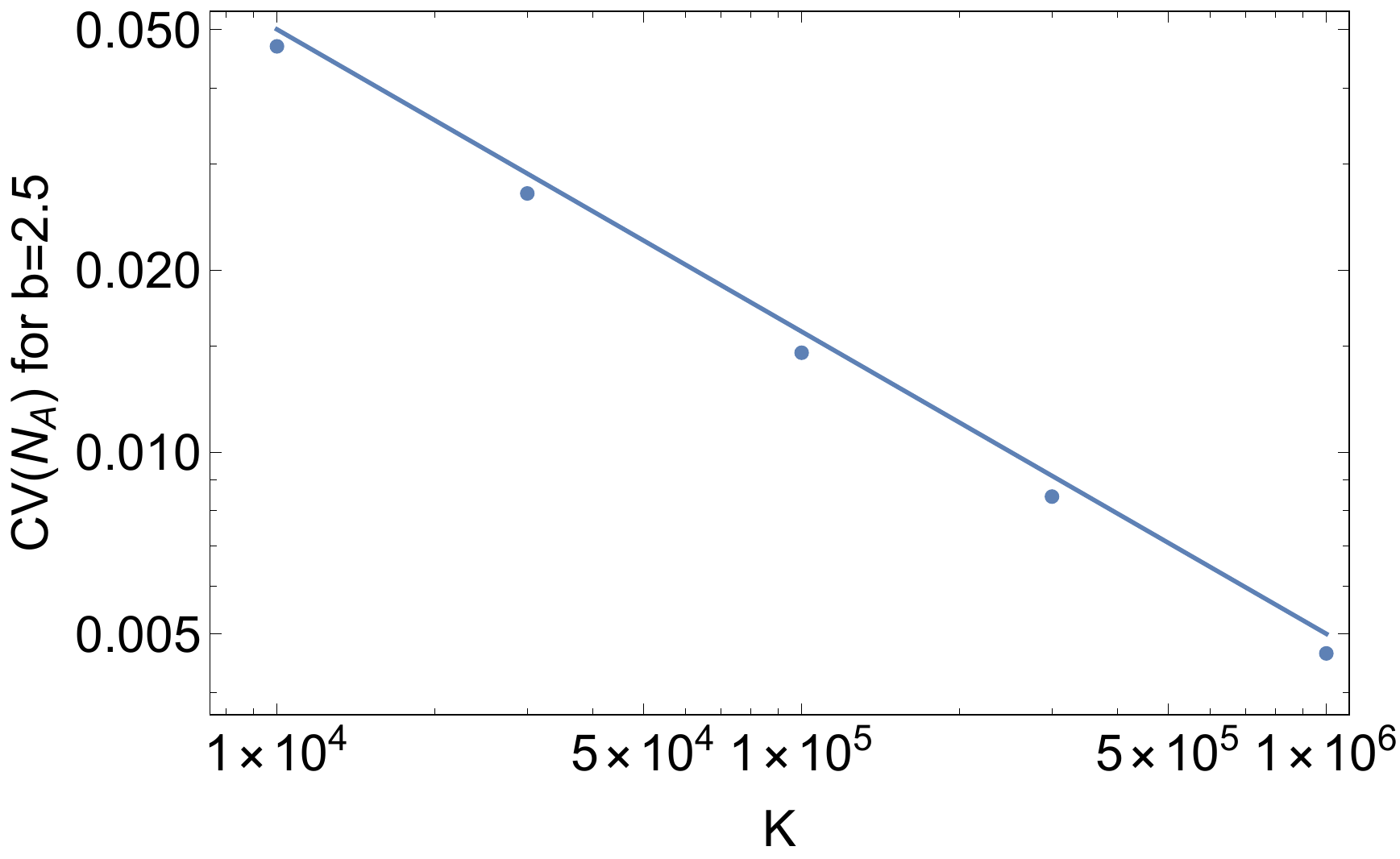}
	\caption{\label{fig:5}Coefficient of variation of $N_A(t)$ for the quasi-synchronous model as a function of $K$, calculated at $b=2.5$ (away from the resonance peak). Solid line represents the scaling $\sim 1/\sqrt{K}$ expected for the Poisson version of the model.}
\end{figure}

The resonance peak becomes sharper with 
increasing carrying capacity $K$ (Fig. \ref{fig:4}, middle). The amplitude of oscillations in the peak is independent of $K$ for a wide range of $K$. This is very different to the scaling $\sim 1/\sqrt{K}$ observed in the Poisson case and also the scaling of CV in the quasi-synchronous model far away from the peak (Fig. \ref{fig:5}). The amplitude of oscillations in the quasi-synchronous model is more than $10$\% of the steady-state population abundance for $K=10^6$, whereas in the Poisson model with identical parameters it is less than 1\%. 

Figure \ref{fig:4}, bottom, shows that the height of the resonance peak decreases with increasing $w$. For $w=0.2$, the peak is barely noticeable. On the other hand, all values $w\leq 0.1$ produce a visible peak.

These results suggest that quasi-synchronous replication leads to a significant enhancement of demographic noise in the quasi-synchronous model. To understand this, let us first revisit what happens in the Poisson version of the model \cite{mckane_predator-prey_2005}. In that model, demographic noise has a flat spectrum and contains a broad range of frequencies (white noise). Frequencies close to the frequency at which the system exhibits damped oscillations are amplified; this leads to quasi-periodic oscillations with the amplitude $\sim \sqrt{K}$. However, since the average abundances $N_A,N_B$ increase proportionally to $K$, the relative magnitude of oscillations decreases as $\sim 1/\sqrt{K}$ with the increasing population size. Noise-induced oscillations are therefore significant only for relatively small systems $K\ll 10^6$. In contrast, here we observe large, persistent oscillations even for $K=10^6$. As we shall see, this can be explained by demographic noise being concentrated in a narrow range of frequencies in the non-Poisson model.

\section{Single-species model}
To understand the spectrum of noise in the quasi-synchronous model, we consider a simpler one-species model. In this model, replication is non-Poissonian with mean doubling time $\ln(2)/b$ as in the two-species model, whereas death is a Poisson process with rate $b N/K$, where $N$ is the total number of cells, $K$ is the carrying capacity, and $b$ is the replication rate. 

In the large-$K$ limit, the average abundance $x=N/K$ evolves according to the logistic equation,
\bq
	dx/dt = bx(1-x).
\eq
The steady-state occupation is $x^*=1$. In the stochastic model ($K<\infty$), the number of cells is thus expected to fluctuate around the mean value $N^*\cong K$.

\begin{figure}
	\includegraphics[width=0.48\columnwidth]{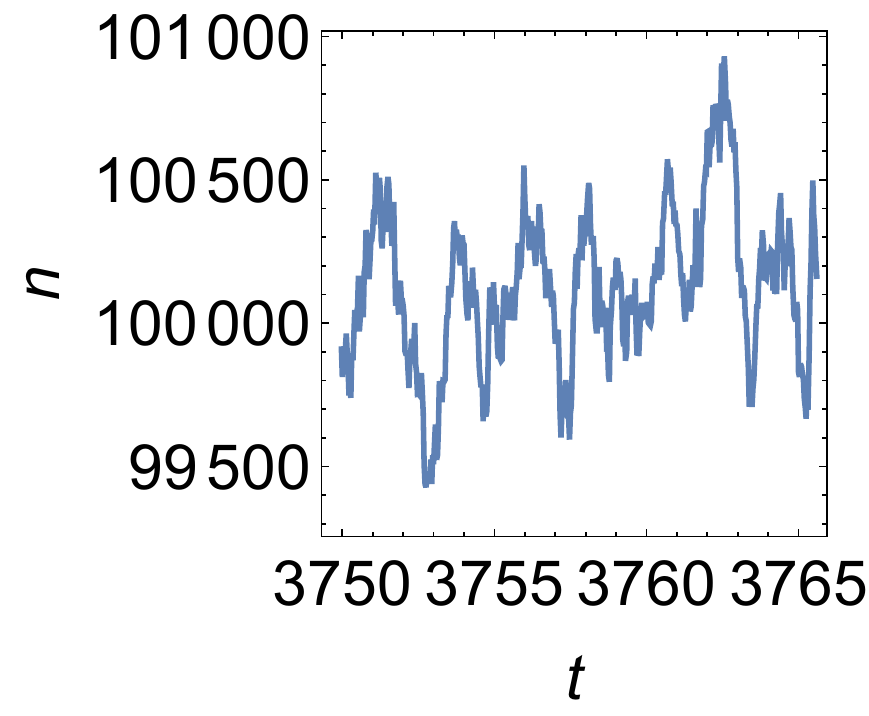}
	\includegraphics[width=0.48\columnwidth]{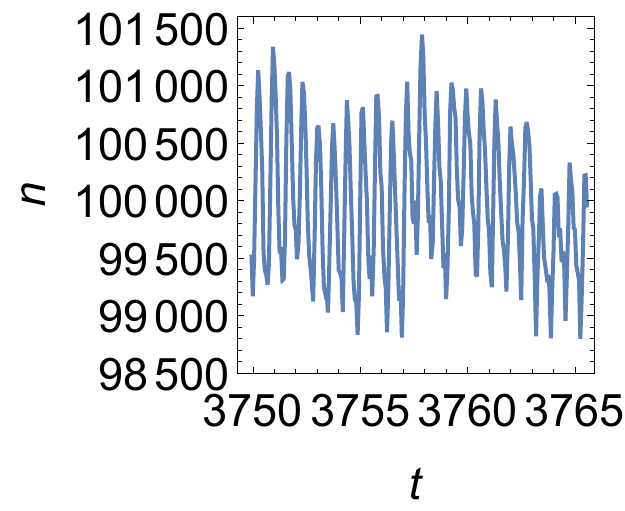}
	\includegraphics[width=0.48\columnwidth]{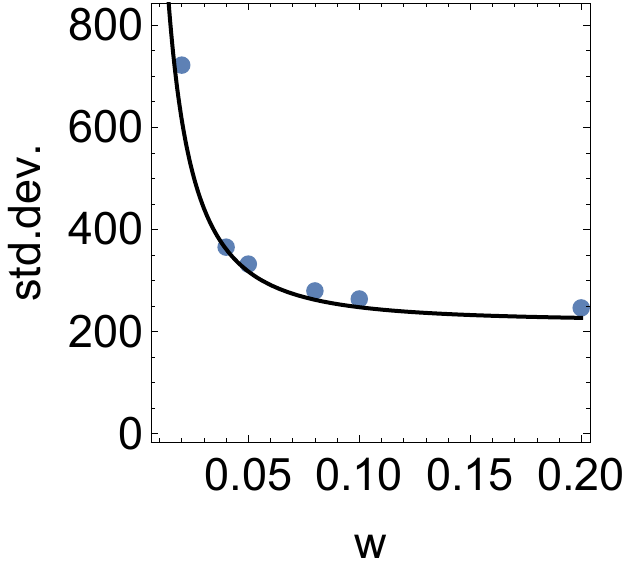}
	\caption{\label{fig:7}Fluctuations in the single-species model. Top: $N(t)$ for $K=100000$: Poisson model (left), and quasi-synchronous model with $w=0.02$ (right). Bottom: standard deviation of fluctuations for different $w$. Points = simulation, line = equation (\ref{eq:dn2}) with  $\gamma = 9.5w^2$. No parameters have been fitted to data here.}
\end{figure}

Figure \ref{fig:7}, top, shows examples of $N(t)$ for the model with Poisson and non-Poisson replication ($w=0.02$), for $K=10^5$. The quasi-synchronous model exhibits more regular oscillations. The standard deviation of $N(t)$ is very similar to the Poisson model for $w>0.05$ but rapidly increases for smaller $w$ (Fig. \ref{fig:7}, bottom).

In what follows, we shall study this model analytically. In particular, we are interested in analytical expressions for (i) the correlation time of oscillations, (ii) the spectrum of oscillations, (iii) the steady-state amplitude of oscillations. This will help us to better understand the behaviour of the two-species model from previous sections.

\subsection{\label{sec1}Preliminary considerations}
We begin by considering the behaviour of a large population of cells in which the cells can be assigned to groups depending on the phase $\phi$ of their cell cycle. The phase is not the same as the timer variable; instead, it should be interpreted as the difference between the timer variable and some arbitrary chosen reference timer. Let $n(\phi)$ be the number density of cells with phase $\phi$. Let us further assume that, if all cells were synchronised (all $\phi$ being equal), the total number of cells would be described by a certain periodic function $f(t)$. This function (besides a different amplitude) also describes number fluctuations in a group of cells that have the same phase $\phi$. The total number of cells in the population is therefore
\bq
  N(t)=\int_{-\infty}^{\infty} n(\phi)f(t-\phi)d\phi,
\eq
which is the convolution of $f$ and $n$. The Fourier spectrum of $N$ is 
\bq
	\mathcal{F}[N](\omega)=\mathcal{F}[f](\omega) \,\mathcal{F}[n](\omega).
\eq
Suppose $n(\phi)$ is Gaussian with variance $\sigma^2$. We have
\bq
\mathcal{F}[N](\omega)=\mathcal{F}[f](\omega) e^{-(1/2)\sigma^2 \omega^2}.
\eq
If $f$ is periodic with angular frequency $\omega_0$, then the lowest-frequency Fourier mode of $N$ at $\omega=\omega_0$ will be reduced in comparison to $f$ by $e^{-(1/2)\sigma^2 \omega^2}$ due to the spread of the phases. All higher modes will be damped even more; we will neglect them for now. For $\omega_0=2\pi/\ln 2$ assumed in our single-species model for $b=1$ (doubling time $\ln 2$) and $\sigma^2\ll 1$, the reduction factor is $e^{[2\pi^2/(\ln 2)^2]\sigma^2}\approx e^{-41.1\sigma^2}$.

\begin{figure}
	\includegraphics[width=0.48\columnwidth]{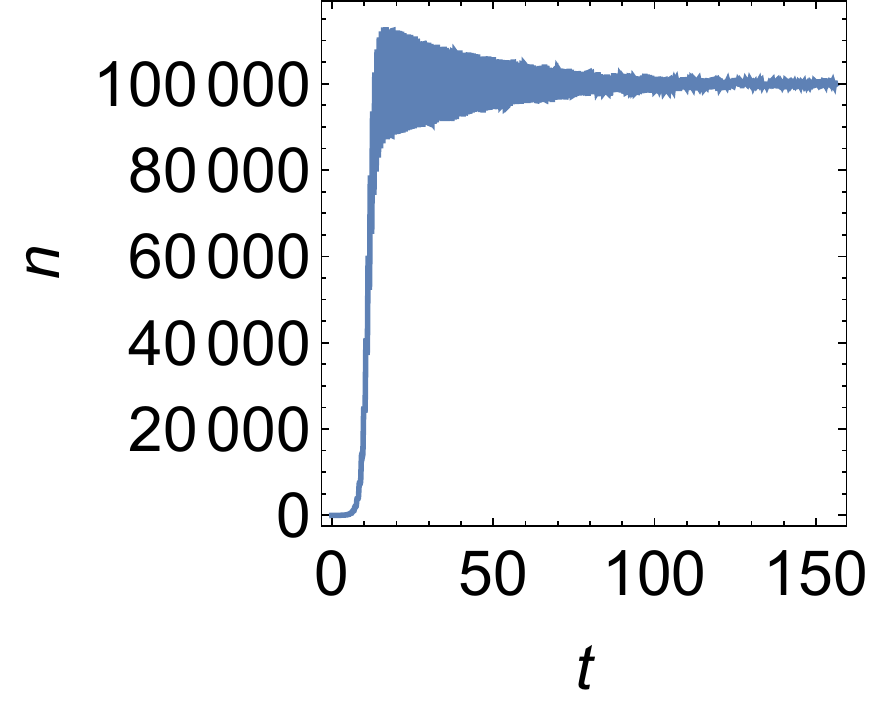}
	\includegraphics[width=0.48\columnwidth]{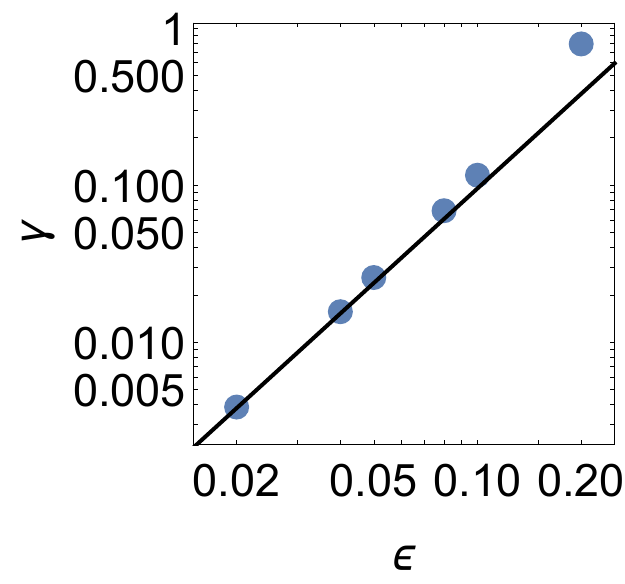}
	\caption{\label{fig:6}Exponential decay of oscillation in the single-species model for $K=100000$. Left: $N(t)$ for $w=0.04$. Right: the exponential decay rate $\gamma$ as a function of $w$ (points). Black line is the theoretical prediction $\gamma\approx 9.5w^2$.}
\end{figure}

Suppose further than each generation causes the distribution $n$ to broaden, due to the finite width of the distribution of doubling times, so that $\sigma^2=\sigma_0^2 (t/ \ln(2))$, where $\sigma_0^2=(w^2/3)(\ln 2)^2$ is the variance of the uniform distribution of doubling times used in the simulations. This corresponds to the variance of the phase distribution increasing by the variance of the doubling time distribution every generation. This will then lead to oscillations in $N(t)$ (caused by quasi-synchronous replication) to decay exponentially with the rate $\gamma =[2\pi^2/(\ln 2)^2](\ln 2)(w^2/3) = [2\pi^2/(3\ln 2)]w^2 \approx 9.5w^2$. Figure \ref{fig:6} shows that the predicted rate is in very good agreement with the decay rate observed in numerical simulations.

\subsection{A more formal approach}
The above result can be derived more formally. We shall start by writing down the equation for the number density of cells $n(\tau,t)$ at time $t$ with the timer variable $\tau$, for the time being neglecting stochastic noise:
\ba
	\partial_t n(\tau,t) = \partial_\tau n(\tau,t) - \frac{\ln 2}{T}n(\tau,t)\int_0^\infty \frac{n(\tau',t)}{K} d\tau' \nonumber \\
	+ 2R(\tau)n(0,t).  \label{eq:single1}
\ea
The first term corresponds to the timer counting backward. The second term represents death with rate proportional to the total size divided by $K$. The factor $(\ln 2)/T$ is required to have the correct behaviour in the limit of perfectly synchronous replication - we shall see this later. The third term represent replication that occurs when the timer reaches $\tau=0$ and is the product of the density of cells $n(0,t)$ in that state and $R(\tau)$, the probability density function for the timer being reset to $\tau$. We assume $R(\tau)$ to be normalized: 
\bq
	\int_0^\infty R(\tau)d\tau =1,
\eq
and that $R(\tau)$ is concentrated around $\tau=T$ as in numerical simulations in previous sections.

\subsection{Stationary solution}
\begin{figure}
	\includegraphics[width=0.48\columnwidth]{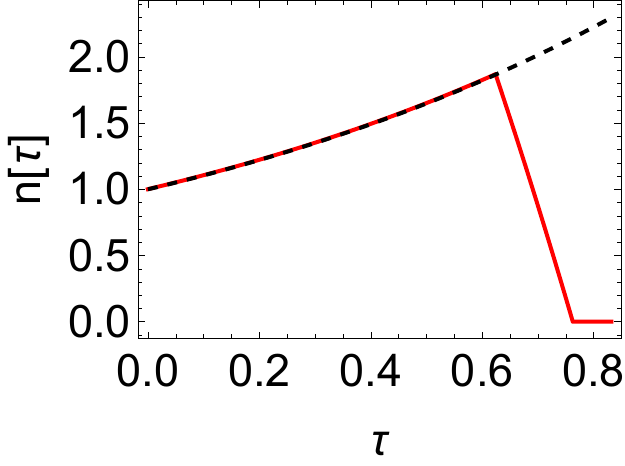}
	\caption{\label{fig:nstar}Stationary cell density $n^*(\tau)$ for $T=\ln 2, w=0.1$. Red line = Eq. (\ref{eq:nstareq}). Black dashed line = approximate solution (\ref{eq:nstarw0}). }
\end{figure}

In the limit $t\to\infty$, Eq. (\ref{eq:single1}) becomes
\bq
	0 = \frac{\partial n(\tau)}{\partial \tau}  -n(\tau) J + 2 n(0) R(\tau),
\eq
where $J=(\ln 2)/T \int_0^\infty (n(\tau)/K) d\tau$, and $n(\tau)$ does not depend on $t$. We can solve this equation for the steady-state distribution $n^*(\tau)$:
\bq
	n^*(\tau)=n^*(0)e^{J\tau}\left[1-2\int_0^\tau e^{-J\tau'} R(\tau') d\tau' \right],
\eq
with the condition $n^*(\tau\to\infty)=0$ implying that
\bq
	\int_0^\infty e^{-J\tau}R(\tau)d\tau = 1/2, \label{eq:J}
\eq
which fixes the value of $J$.
If $R$ is a uniform distribution with mean $\tau=T$ and width $2w T$, we obtain from (\ref{eq:J}) that
\bq
	e^{-JT} \frac{\sinh(JTw)}{JTw} = 1/2.
\eq
This equation must be solved for $J$ numerically. In addition, one must determine the value of $n^*(0)$ from the relationship between $J$ and $n(\tau)$:
\bq
	JK=n^*(0) \int_0^{T(1+w)} e^{J\tau}\left[1-2\int_0^\tau e^{-J\tau'} R(\tau') d\tau' \right] d\tau \label{eq:nstareq}
\eq
Figure \ref{fig:nstar} shows an example of $n^*(\tau)$ for $w=0.1$, calculated in this way. The cell number density is proportional to $e^{J\tau} \approx 2^{\tau/T}$ for $\tau<T$, and rapidly falls down to zero for $\tau>T$.
The solution simplifies greatly in the limit $w\to 0$, in which $J$ tends to $(\ln 2)/T$. The steady state number density becomes then
\bq
  n^*(\tau) = K\frac{\ln 2}{T} 2^{\tau/T} \label{eq:nstarw0}
\eq

\subsection{\label{sec:pert}Evolution of a small perturbation}
We now consider the time evolution of a small perturbation to the steady state solution:
\bq
	n(\tau,t) = n^*(\tau)(1+\epsilon(\tau,t)).
\eq
of the noise-less equation \ref{eq:single1}. Inserting this into Eq. \ref{eq:single1} gives:
\ba
n^* \partial_t{\epsilon} &=& (\partial_\tau n^*)(1+\epsilon) + n^* \partial_\tau\epsilon \nonumber \\ 
&-& \frac{\ln 2}{TK}n^* (1+\epsilon)\int_0^\infty n^*(\tau')(1+\epsilon(\tau',t))d\tau' \nonumber \\
& +& 2R(\tau)n^*(1+\epsilon) . \nonumber \\
\ea
We note that $\partial_\tau n^*=Jn^* - 2n^*(0)R(\tau)$, and keep only terms linear in $\epsilon$:
\ba
n^* \partial_t{\epsilon} &=& n^*\partial_\tau\epsilon - \frac{\ln 2}{TK} n^* \int_0^T n^*(\tau)\epsilon(\tau',t)d\tau' \nonumber \\ &+&2R(\tau)n^*(0)(\epsilon(0,t)-\epsilon(\tau,t)) .
\ea
We divide by $n^*$ and obtain
\ba
\partial_t{\epsilon} &=& \partial_\tau\epsilon - \frac{(\ln 2)}{TK} \int_0^\infty n^*(\tau')\epsilon(\tau',t)d\tau' \nonumber \\
&-& 2\frac{n^*(0)}{n^*(\tau)}R(\tau)(\epsilon(0,t)-\epsilon(\tau,t))
\ea
with the boundary condition $\epsilon(0,t)=\epsilon(T,t)$.
We now expand $\epsilon$ and $\frac{n^*(0)}{n^*(\tau)}R(\tau)$ as Fourier series (consistent with the b.c.):
\ba
\epsilon(\tau,t) = \sum_{k=-\infty}^\infty A_k(t) e^{2\pi i k  \tau/T} , \\
\frac{n^*(0)}{n^*(\tau)}R(\tau) = \sum_{k=-\infty}^\infty R_k(t) e^{2\pi i k \tau/T} .\label{eq:nrffnonoise}	
\ea
where the coefficients $\{R_k\}$ are given by
\bq
	R_k = \frac{1}{T} \int_0^T e^{-2\pi i k \tau/T} \frac{n^*(0)}{n^*(\tau)}R(\tau) d\tau . \label{eq:Rks}
\eq
The transformed equation reads
\ba
& &\sum_k \partial_t{A}_k e^{2\pi i k \tau/T} = \sum_k A_k(2\pi i k/T)e^{2\pi i k \tau/T} - \nonumber \\
& &- \frac{(\ln 2)}{TK} \sum_k A_k \int_0^T n^*(\tau')e^{2\pi i k \tau'/T}d\tau'  \nonumber \\
& &+ 2 \sum_m R_m e^{2\pi i m \tau/T} \left(\sum_k A_k -\sum_k A_k e^{2\pi i k \tau/T}\right). \label{eq:ffknonoise} 
\ea
The sums in (\ref{eq:ffknonoise}) can be compared term by term since they must be valid for any $\tau$. 
This leads to the following equation for the Fourier coefficients $A_k(t)$:
\ba
\partial_t{A}_k &=& (2\pi i k/T)A_k \nonumber \\
&-& \delta_{k,0}\frac{\ln 2}{TK}\sum_m A_m \int_0^T n^*(\tau')e^{2\pi i m \tau'/T}d\tau' \nonumber \\
&+& 2(R_k \sum_m A_m - \sum_m R_m A_{k-m}). 
\ea
In particular, for $k>0$ we have
\bq
\partial_t{A}_k = (2\pi i k/T) A_k + 2(R_k \sum_m A_m - \sum_m R_m A_{k-m}).
\eq
Let us assume that the initial perturbation is a pure $k$th Fourier mode, i.e., $A_k\neq 0$ only for a single value of $k$.
The first term represents oscillations with period $T/k$ of that mode, which essentially gives a travelling-wave type of solution $\sim\exp(2\pi i k(\tau-t)/T)$. The second term represents damping with rate $\gamma_k = -2(R_k - R_0)$. We can calculate this rate using equation (\ref{eq:Rks}) in the limit $w\to 0$, since then we have from Eq. (\ref{eq:nstarw0}) that $n^*(0)/n^*(\tau)=2^{-\tau/T}$ and hence
\bq
	R_k = \frac{1}{T}\int_{T(1-w)}^{T(1+w)} e^{-2\pi i k \tau/T} 2^{-\tau/t} \frac{1}{2Tw}d\tau.
\eq
We obtain that
\bq
	\gamma_k= -2(R_k - R_0) \cong \frac{2\pi^2}{3T}k^2w^2,
\eq
which, for $T=\ln 2$ and $k=1$ reproduces the decay rate $\gamma\approx 9.5w^2$ which we have already seen in Sec. \ref{sec1}.

\subsection{\label{sec:synosc}Amplitude of oscillations for perfectly synchronous replication}
We shall now add noise to the model and see how it affects its behaviour. We shall first consider a fully synchronous replication with arbitrary period $T$, which leads to the following equation:
\bq
	\partial_t n = \partial_\tau n -  \frac{\ln 2}{T}n\int_0^T \frac{n(\tau',t)}{K}d\tau' + \sqrt{n^*}\eta , \label{eq:dndt}
\eq
with boundary conditions
\bq
n(0,t)=(1/2)n(T,t).
\eq
The noise term $\sqrt{n^*}\eta$ is due to death only, since replication is perfectly synchronous. We assume $\eta(\tau,t)$ represents uncorrelated white noise: 
\bq
	\left<\eta(\tau_1,t_1)\eta(\tau_2,t_2)\right> = D\delta(\tau_1-\tau_2)\delta(t_1-t_2), \label{eq:noisedef}
\eq
with some $D>0$ to be specified later. It can be easily verified that this form of noise arises from a master equation for the model with no replication by performing a van Kampen expansion \cite{gardiner_stochastic_2009} of the master equation. While it may be possible to derive the noise term also in the presence of non-Markovian replication, we find it easier to postulate that Eq. (\ref{eq:noisedef}) generally holds, and justify it based on the agreement between the result of our calculation and the computer simulation (see below).

In the absence of noise, equation (\ref{eq:dndt}) has the steady-state solution
\ba
	n^*(\tau) = K\frac{\ln 2}{T}2^{\tau/T}, \\
	\int_0^T n^*(\tau)d\tau = K . \label{eq:nstartot}
\ea
To solve the time-dependent equation with noise, we consider a small perturbation (similarly as in the previous section):
\bq
	n(\tau,t) = n^*(\tau)(1+\epsilon(\tau,t)).
\eq
This gives
\ba
	n^* \partial_t{\epsilon} &=& (\partial_\tau n^*)(1+\epsilon) + n^* \partial_\tau\epsilon \nonumber \\ 
	&-& \frac{\ln 2}{TK}n^* (1+\epsilon)\int_0^T n^*(\tau')(1+\epsilon(\tau',t))d\tau' + \sqrt{n^*}\eta . \nonumber \\
	& &
\ea
We note that $\partial_\tau n^*=((\ln 2)/T)n^*$, and only keep terms linear in $\epsilon$:
\bq
	n^* \partial_t{\epsilon} = n^*\partial_\tau\epsilon - \frac{(\ln 2)^2}{T^2} n^* \int_0^T 2^{\tau'/T}\epsilon(\tau',t)d\tau' + \sqrt{n^*}\eta .
\eq
We divide by $n^*$ and obtain
\bq
	\partial_t{\epsilon} = \partial_\tau\epsilon - \frac{(\ln 2)^2}{T^2} \int_0^T 2^{\tau'/T}\epsilon(\tau',t)d\tau' + (n^*)^{-1/2}\eta ,
\eq
with the following boundary and initial conditions: $\epsilon(0,t)=\epsilon(T,t)$ and $\epsilon(\tau,0)=0$.
Proceeding as in the previous section, we expand $\epsilon$ and $(n^*)^{-1/2}\eta$ as Fourier series (consistent with the b.c.):
\ba
	\epsilon(\tau,t) = \sum_{k=-\infty}^\infty A_k(t) e^{2\pi k i \tau/T} , \\
	(n^*(\tau))^{-1/2}\eta(\tau,t) = \sum_{k=-\infty}^\infty \eta_k(t) e^{2\pi k i \tau/T} .\label{eq:netaff}\\	
\ea
The transformed equation reads
\ba
	& &\sum_k \partial_t{A}_k e^{2\pi k i \tau/T} = \sum_k A_k(2\pi i k/T)e^{2\pi k i \tau/T} - \nonumber \\
	& &- \frac{(\ln 2)^2}{T^2} \sum_k A_k \int_0^T 2^{\tau'/T}e^{2\pi k i \tau'/T}d\tau'  \nonumber \\
	& &+ \sum_k \eta_k(t) e^{2\pi k i \tau/T} . \label{eq:ffk} \nonumber
\ea
The integral over $d\tau'$ gives
\bq
	\int_0^T 2^{\tau'/T}e^{2\pi k i \tau'/T}d\tau' = \frac{iT}{i\ln 2-2\pi k}.
\eq
Comparing the sums in (\ref{eq:ffk}) term-by-term we notice that the $(\ln 2)^2$ term does not contain any factor $e^{2\pi k i \tau/T}$, so it only contributes to the constant term:
\bq
	\delta_{k,0} \frac{(\ln 2)^2}{T} \sum_n A_n \frac{i}{i\ln 2-2\pi n} \approx \delta_{k,0}\frac{\ln 2}{T}A_0,
\eq
where we have assumed that all $A_n$ for $n\neq 0$ are much smaller than $A_0$ (we shall see later that this is the case). This leads to the following equation for the Fourier coefficients $A_k(t)$:
\bq
	\partial_t{A}_k = (2\pi i k/T)A_k - ((\ln 2)/T)\delta_{k,0}A_0 + \eta_k . \label{eq:Ak}
\eq
In particular, for $k=0$ we have
\bq
	\partial_t{A}_0 = -\frac{\ln 2}{T}A_0 + \eta_0 ,
\eq
which can be formally solved as
\bq
	A_0(t) = 2^{-t/T} \int_0^t 2^{t'/T}\eta_0(t')dt' .
\eq
This gives
\bq
	\left<A_0A_0^\dagger\right>(t) = 2^{-\frac{2t}{T}}\int_0^t dt_1 \int_0^t dt_2 2^{\frac{t_1+t_2}{2}} \left<\eta_0(t_1)\eta_0^\dagger(t_2)\right> .
\eq
Equation (\ref{eq:netaff}) enables us to write
\bq
	\eta_k(t) = \frac{1}{T}
	\int_0^T \left(\frac{K\ln 2}{T}\right)^{-1/2} 2^{-\frac{\tau}{2T}} \eta(\tau,t) e^{-2\pi i k \tau/T}  d\tau.
\eq
The average of the noise term gives
\ba
	&&\left<\eta_k(t_1)\eta_k^\dagger(t_2)\right> = \frac{1}{TK\ln 2}\times \nonumber \\
	&&\times \int_0^T d\tau_1 \int_0^T d\tau_2 2^{-\frac{\tau_1+\tau_2}{2T}} e^{-\frac{2\pi i k (\tau_1-\tau_2)}{T}}	\left<\eta(\tau_1,t_1)\eta^\dagger(\tau_2,t_2)\right> \nonumber \\
	&& = \frac{D}{K}\frac{\delta(t_1-t_2)}{2(\ln 2)^2} .
\ea
We therefore have
\ba
	\left<A_0A_0^\dagger\right>(t) &=& 2^{-2t/T} \int_0^t  \frac{2^{2t'/T}D}{K2(\ln 2)^2	}  dt' \nonumber \\
	&=& \frac{DT}{K}\frac{1-2^{-2t/T}}{4(\ln 2)^3 }.
\ea
Proceeding similarly for $k\neq 0$, we obtain:
\bq
	A_k(t) = e^{2\pi i k t/T}\int_0^t e^{-2\pi i k t'/T}\eta_k(t')dt',
\eq
from which we obtain that
\ba
	\left<A_kA_k^\dagger\right>(t) &=& \int_0^t dt_1 \int_0^t dt_2 e^{-2\pi ik(t_1-t_2)/T} \left<\eta_k(t_1)\eta_k^\dagger(t_2)\right> \nonumber \\
	&=& \frac{D}{K}\frac{t}{2(\ln 2)^2}.
\ea
We can now calculate the standard deviation of $\Delta N$, the difference between the total number of cells at time $t$ and the average steady-state number:
\ba
	\Delta N(t) &=& \int_0^T n^*(\tau)\epsilon(\tau,t)d\tau \\
	&=& \int_0^T \frac{K\ln 2}{T}2^{\tau/T} \sum_k A_k(t) e^{2\pi i k \tau/T} d\tau \nonumber \\
	&=& \sum_k A_k(t) \frac{K\ln 2}{T} \int_0^T 2^{\tau/T} e^{2\pi i k \tau/T} d\tau \nonumber \\
	&=& (K\ln 2) \sum_k A_k(t) \frac{i}{i\ln 2-2\pi k} .
\ea
This gives (we note that terms $\left<A_k A_n^\dagger\right>$ with $k\neq n$ vanish):
\ba
	\left<|\Delta N(t)|^2\right> & = &(K\ln 2)^2 \sum_{k=-\infty}^\infty  \frac{\left<A_kA_k^\dagger\right>(t)}{(\ln 2)^2+(2\pi k)^2} \nonumber \\
	= &KD&\left[T\frac{1-2^{-2t/T}}{4(\ln 2)^3 } + 2\sum_{k=1}^\infty \frac{t/2}{(\ln 2)^2+(2\pi k)^2} \right] \nonumber \\
	= &KD&\left[T\frac{1-2^{-2t/T}}{4(\ln 2)^3} + t\frac{3(\ln 2)-2}{4(\ln 2)^2}\right] . \label{eq:dn2t}
\ea
In the limit $t\to\infty$ this gives
\bq
	\left<|\Delta N(t)|^2\right> \cong KDt\frac{3(\ln 2)-2}{4(\ln 2)^2} \approx 0.0413 \, KDt . 
	\label{eq:vardN2}
\eq
Equation (\ref{eq:vardN2}) predicts that the amplitude of oscillations for perfectly synchronous replication increases linearly in time. However, recall that our result has been derived under the assumption of a small perturbation. In reality, the amplitude will be limited by non-linear effects. 

\subsection{Amplitude of quasi-synchronous oscillations}
Let us now consider the case of quasi-synchronous replication. Rather than attempting to solve Eq. (\ref{eq:dndt}) with the extra term $R(\tau)$ as in Eq. (\ref{eq:single1}), we observe (as argued in subsection \ref{sec:pert}) that the $k$-th Fourier mode will be damped with rate $k^2\gamma$. We thus consider the following modification to Eq. (\ref{eq:Ak}) for $k\neq 0$:
\bq
	\partial_t{A}_k=(2\pi i k/T - k^2\gamma)A_k + \eta_k ,
\eq
where $\gamma$ is the damping coefficient derived previously.
The equation for $k=0$ remains unchanged. Proceeding as in Sec. \ref{sec:synosc}, we obtain
\bq
	\left<A_kA_k^\dagger\right>(t) = \frac{D}{K} \frac{1-e^{-2k^2\gamma t}}{4(\ln 2)^2 k^2 \gamma} .
\eq
Inserting this into the equation for $\left<|\Delta N(t)|^2\right>$ we have in the limit $t\to\infty$:
\ba
	& &\left<|\Delta N|^2\right> = \left<|\Delta N(t\to\infty)|^2\right> 
	= \nonumber \\
	&=& KD\left[\frac{T}{4(\ln 2)^3} + 2\sum_{k=1}^\infty \frac{1}{4k^2\gamma}\frac{1}{(\ln 2)^2+(2\pi k)^2} \right] \nonumber \\
	&=& KD\left[\frac{T}{4(\ln 2)^3} + \pi^2\frac{12-18\ln 2+(\ln 2)^2}{12\gamma(\ln 2)^4}\right]  .
\ea
For $T=\ln 2$ we obtain that
\bq
	\left<|\Delta N|^2\right> \approx KD(0.5203+0.01355/\gamma).
	\label{eq:dn2}
\eq
It remains to relate $D$ to the parameters of the model. We again assume that death is the main source of stochasticity, and that the contribution from quasi-synchronous replication is negligible. Consider a pure death process with the same total number of organisms $K$ as the steady state total (\ref{eq:nstartot}), and death rate $d=(\ln 2)/T$ as per Eq. (\ref{eq:dndt}). For short time intervals, the variance of the number of organisms in the pure death process equals to $\left<|\Delta N|^2\right>=(Kd)t=(K(\ln 2)/T)t$ (easy to derive from the general formula on p. 108-109 of Ref. \cite{athreya_branching_2004}). 
On the other hand, from Eq. (\ref{eq:dn2t}) we have that for small $t$ and $T=\ln 2$,
\bq
	\left<|\Delta N|^2\right> \cong KD \frac{3}{4\ln 2} t.
\eq
Comparing the two formulas for $\left<|\Delta N|^2\right>$, we obtain that
\bq
	D=\frac{4\ln 2}{3} \approx 0.924.
\eq
Figure \ref{fig:7} shows that equation (\ref{eq:dn2}) with the above value of $D$ reproduces the variance from numerical simulations very well. 

\subsection{\label{sec:sp1}Spectrum of fluctuations}
\begin{figure}
	\includegraphics[width=0.98\columnwidth]{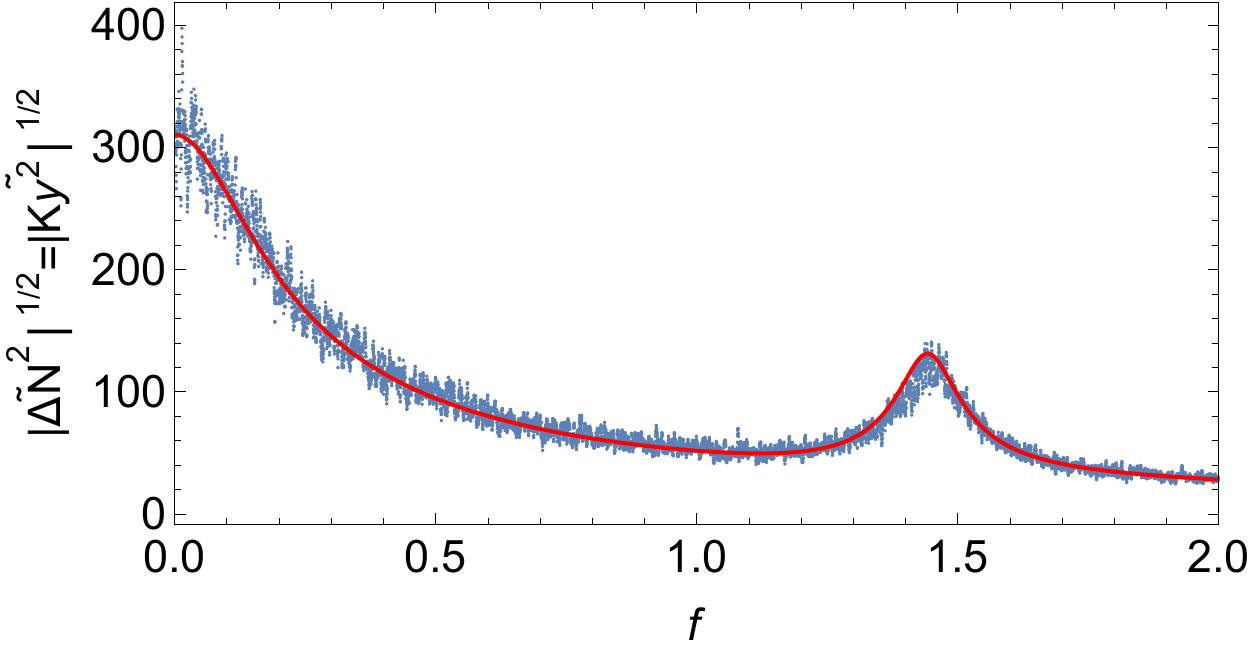}
	\includegraphics[width=0.98\columnwidth]{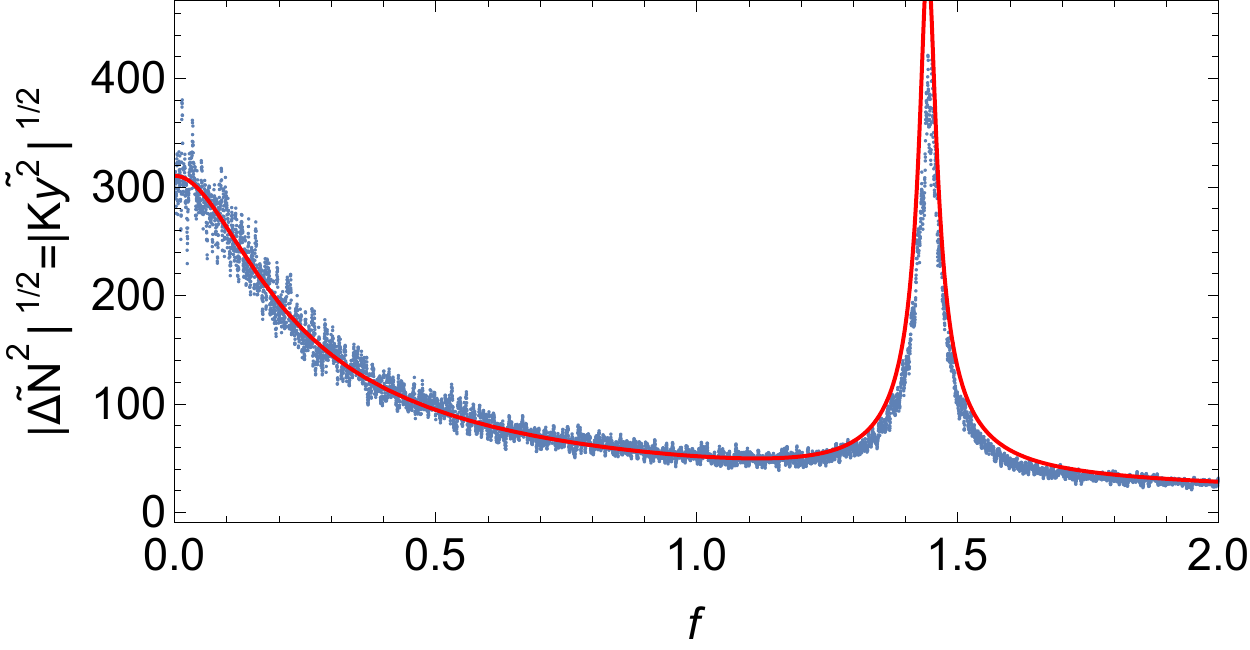}
	\includegraphics[width=0.98\columnwidth]{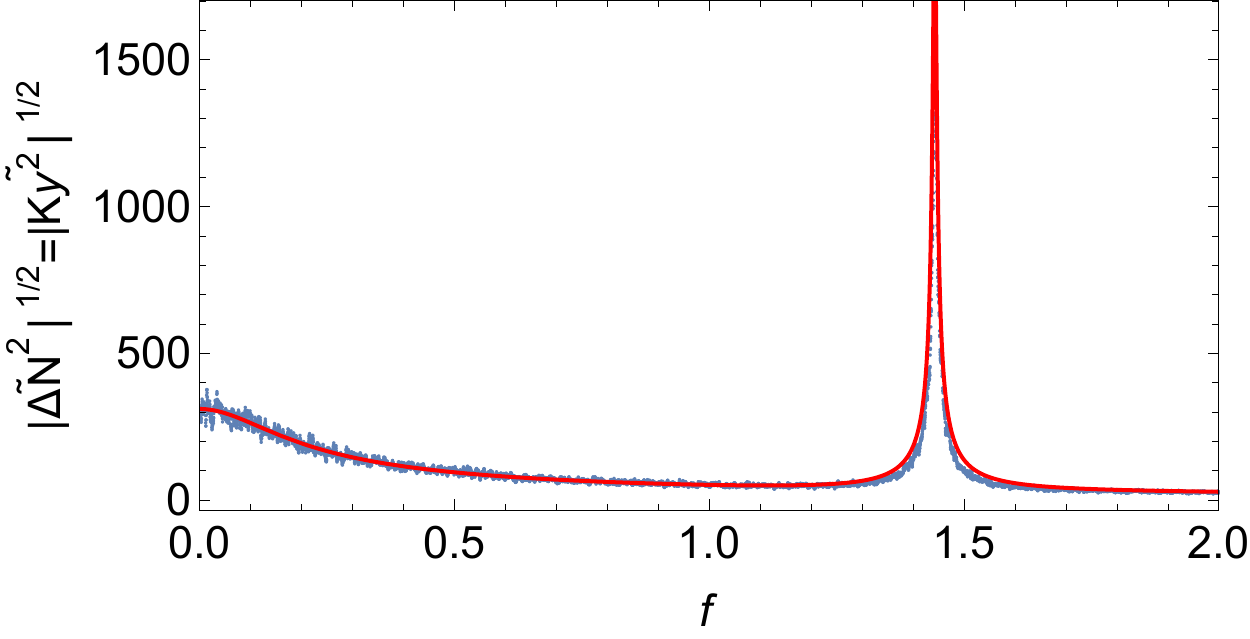}
	\caption{\label{fig:8}Fourier spectrum $|K\tilde{y}^2|^{1/2}$ of $N(t)$ in the single-species model for $K=10^5$. Blue points = simulations, red lines = theoretical prediction (no fitting) obtained from Eq. (\ref{eq:yfftfinalf}). From top to bottom: $w=0.2,0.1,0.05$. }
\end{figure}

We can now obtain a very good analytic approximation for the spectrum of normalized fluctuations $y(t)=N(t)/K-1$ in the single-species model by Fourier-transforming the expression for $\Delta N$:
\bq
	\tilde{y} (\omega) = (\ln 2)\sum_k \tilde{A}_k(\omega)\frac{i}{i\ln 2 - 2\pi k} ,
\eq
in which
\bq
	\tilde{A}_k(\omega) = \lim_{L\to\infty} \frac{1}{\sqrt{L}} \int_{-L/2}^{L/2} A_k(t) e^{i \omega t} dt . \label{eq:Akft}
\eq
We have
\ba
	\left<|\tilde{A}_0(\omega)|^2\right> = \frac{\left<|\tilde{\eta}_0|^2\right>}{\left(\frac{\ln 2}{T}\right)^2 + \omega^2} , \\
	\left<|\tilde{A}_k(\omega)|^2\right> = \frac{\left<|\tilde{\eta}_k|^2\right>}{(\omega T-2\pi k)^2+\gamma^2 T^2 k^4} .
\ea
in which $\left<|\tilde{\eta}_k|^2\right>$ is defined through the Fourier transform like in Eq. (\ref{eq:Akft}), and evaluates to
\bq
	\left<|\tilde{\eta}_0|^2\right>=\left<|\tilde{\eta}_k|^2\right> = \frac{D}{K 2(\ln 2)^2} = D_2 K^{-1},
\eq
with $D_2=\frac{2}{3\ln 2}\approx 0.962$. This gives
\ba
	& &K\left<|\tilde{y}|^2(\omega)\right> = (\ln 2)^2 \sum_k \frac{\left<|\tilde{A}_k|^2\right>(\omega)}{(\ln 2)^2+(2\pi k)^2} \nonumber \\
	& = & \frac{D_2}{\left(\frac{\ln 2}{T}\right)^2 + \omega^2} + \nonumber \\
	& +& \sum_{k=1}^\infty \frac{2D_2T^2}{(\omega T-2\pi k)^2+\gamma^2 T^2 k^4} \frac{(\ln 2)^2}{(\ln 2)^2+(2\pi k)^2}
	\nonumber \\
	& \approx & \frac{D_2}{\left(\frac{\ln 2}{T}\right)^2 + \omega^2} + \nonumber \\
	& +& \frac{2D_2T^2(\ln 2)^2}{[(\omega T-2\pi)^2+\gamma^2 T^2][(\ln 2)^2+(2\pi)^2]} + \dots
	\label{eq:yfftfinalT}
\ea
where `$\dots$' stand for terms corresponding to higher harmonics which we neglect because they are strongly suppressed by the denominator increasing fast with $k$. 
For $T=\ln 2$, we have 
\bq
	K\left<|\tilde{y}|^2(\omega)\right> \cong \frac{D_2}{1+\omega^2} + \frac{2(\ln 2)^2D_2}{(\ln 2)^2+(2\pi)^2} \frac{1}{(\omega-\frac{2\pi}{\ln 2})^2+\gamma^2}. \label{eq:yfftfinal}
\eq
The formula as a function of frequency $f=\omega/(2\pi)$ reads:
\ba
K\left<|\tilde{y}|^2(f)\right> \cong \nonumber \\
\cong  \frac{D_2/(2\pi)^2}{1+(2\pi f)^2} 
+ \frac{D_2\frac{2(\ln 2)^2}{(2\pi)^2}}{(\ln 2)^2+(2\pi)^2} \frac{1}{(2\pi f-\frac{2\pi}{\ln 2})^2+\gamma^2}, \label{eq:yfftfinalf}
\ea
where the factor $1/(2\pi)^2$ is required for correct normalization.
Figure \ref{fig:8} shows that equation (\ref{eq:yfftfinalf}) agrees well with the numerically obtained spectrum for a broad range of $w$ values. 

Now we have determined the spectrum of the single-species model, we can proceed to obtain the spectrum of the two-species model.

\section{Approximate analytic solution of the two-species model}
\subsection{Linear approximation}
We shall assume that near the steady state the dynamics of the two-species model can be described by linearized equations
\ba
	dy_A/dt &=& a_{AB}y_B + \eta_A , \label{eq:dyA} \\
	dy_B/dt &=& a_{BA}y_A + a_{BB} y_B + \eta_B , \label{eq:dyB} 
\ea
where $y_A=x_A-x_A^*, y_B=x_B-x_B^*$ ($x_A^*,x_B^*$ are steady-state concentrations), and $a_{AB},a_{BA},a_{BB}$ are given by the following expressions
\ba
a_{AB} &=& \frac{p_3 (b-p_2)+p_1(b+p_3)}{p_4}, \\
a_{BB} &=& \frac{p_3 (b-p_2)}{p_1}, \\
a_{BA} &=& \frac{p_4 (b-p_2)}{p_1}.
\ea
Here $\eta_A,\eta_B$ represent noise (not necessarily white noise) due to replication and death of both species.

\begin{figure}
	\includegraphics[width=8cm]{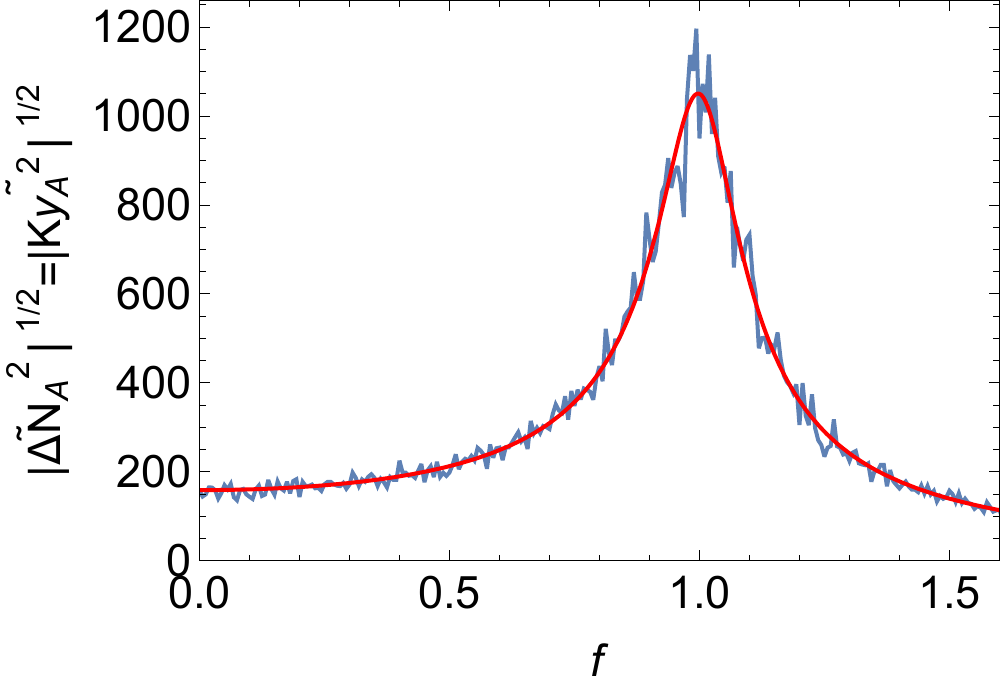}
	\caption{\label{fig:xapoisson}Plot of the spectrum of $y_A$ versus the frequency $f=\omega/(2\pi)$, for the Poisson model. Blue = simulation with $K=10^5,b=1$, and the remaining parameters as in $S_{0.5}$. Red = analytic expression (\ref{eq:xAspec}) with $\left<|\eta_A|^2\right>=\left<|\eta_B|^2\right>=bK$, and $\left<|\eta_{AB}|^2\right>=0$. }
\end{figure}

Fourier-transforming Eqs. (\ref{eq:dyA}, \ref{eq:dyB}) leads to the following expression for the spectrum of $y_A$:
\begin{widetext}
\bq
	\left<|\tilde{y}_A(\omega)|^2\right> = 	  \frac{(a_{BB}^2+\omega^2) \left<|\eta_A|^2\right> -2 a_{AB} a_{BB} \left<|\eta_A\eta_B|^2\right> + a_{AB}^2\left<|\eta_B|^2\right>}{a_{AB}^2 a_{BA}^2+2 a_{AB} a_{BA} \omega ^2+a_{BB}^2 \omega ^2+\omega ^4} .
\label{eq:xAspec}
\eq
\end{widetext}
The formula for the spectrum of $y_B$ (not shown) is very similar.
Figure \ref{fig:xapoisson} shows that Eq. (\ref{eq:xAspec}) works well for the Poisson case (asynchronous replication), for which we assume $\left<|\eta_A|^2\right>=\left<|\eta_B|^2\right>=(2b)(K/2)=bK$ (recall that $K/2$ is the average number of organisms for our choice of the parameters, and the factor $2b$ is due to both birth and death contributing equally near the steady state), and $\left<|\eta_{AB}|^2\right>=0$. 

We now want to establish whether Eq. (\ref{eq:xAspec}) also works for the non-Poissonian case, with an appropriate choise of the coloured noise $\eta_A,\eta_B$, based on the single-species calculation presented in the previous Sec. \ref{sec:sp1}. We assume that fluctuations around the steady state in the single species model can effectively be described by the following equation:
\bq
	dy/dt = -by + {\rm coloured\,noise}.
\eq
Since we know the spectrum of $y$, we can calculate the spectrum of the coloured noise as $dy/dt$ or, in Fourier space, by multiplying Eq. (\ref{eq:yfftfinal}) by $\omega^2$:
\ba
\left<|\tilde{\eta}_{A,B}|^2(\omega)\right> = K_{A,B}^{-1}\omega^2 \left(\frac{D_2}{b^2+\omega^2} + \right. \nonumber \\
\left. + \frac{2(\ln 2)^2 D_2}{(\ln 2)^2+(2\pi)^2} \frac{1}{(\omega-\frac{2\pi}{T})^2+\gamma^2} \right), \label{eq:etafftfinal}
\ea
with $K_A=Kx_A^*,K_B=Kx_B^*$. We then insert Eq. (\ref{eq:etafftfinal}) into Eq. (\ref{eq:xAspec}), assuming again that $\left<|\eta_{AB}|^2\right>=0$ (justified since both species replicate independently with rates unaffected by the other species).

\begin{figure}
	\includegraphics[width=8cm]{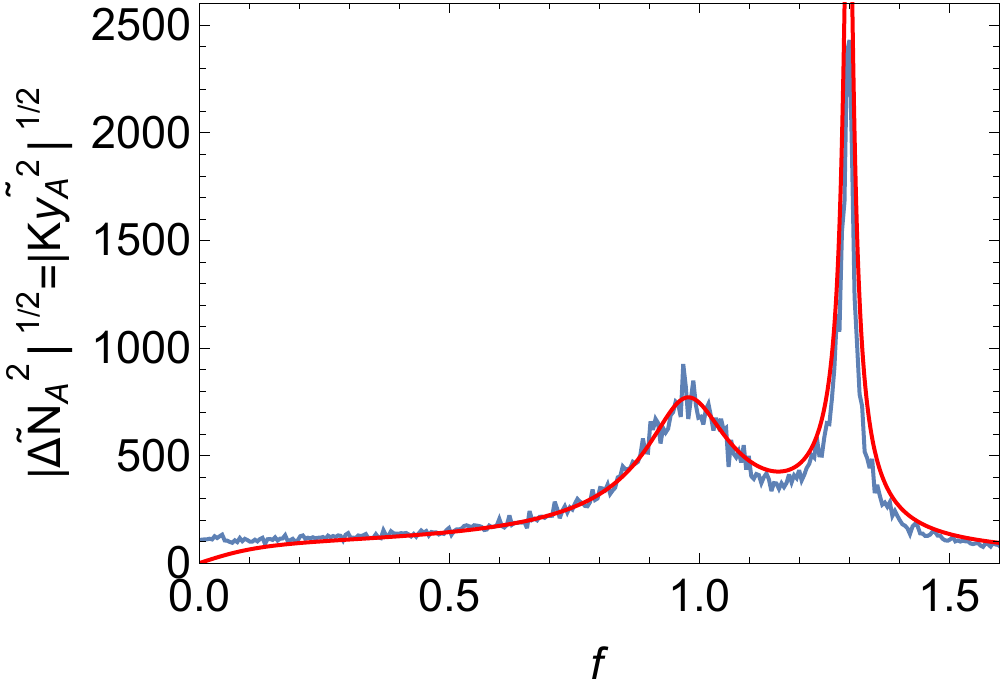}
	\includegraphics[width=8cm]{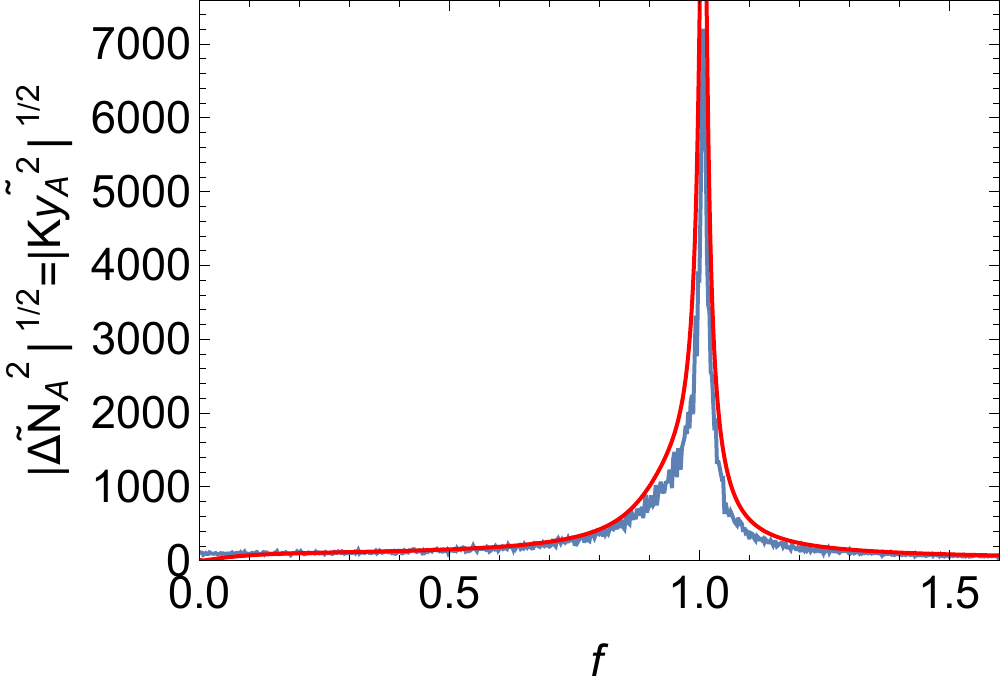}
	\caption{\label{fig:xadetuned}Plot of the spectrum of $y_A$ versus frequency $f=\omega/(2\pi)$, for the non-Poisson model with the replication frequency slightly detuned ($b=0.9$, top) and in resonance ($b=0.7$, bottom). Blue = simulation with $K=10^5, w=0.08$, and the remaining parameters as in $S_{0.5}$. Red = analytic expression \ref{eq:xAspec} with the noise terms from Eq. (\ref{eq:etafftfinal}), and $\gamma=6.579w^2$ (here $T=1$). }
\end{figure}

Figure \ref{fig:xadetuned}, shows that this simple approach works quite well for different frequencies of replication (controlled by $b$). If the replication frequency is slightly detuned from the natural frequency of the system (Fig. \ref{fig:xadetuned}, top), two peaks are visible in the spectrum: a sharp peak coming from the quasi-synchronous birth events, and a much wider but lower peak corresponding to white noise-induced oscillations at the natural frequency $f\approx 1$. 

Figure \ref{fig:xadetuned}, bottom, shows that when $b=0.7$ is tuned in to the resonant frequency of the system, only one peak is visible, with a slight broadening towards lower frequencies.

\subsection{Amplitude of oscillations}
We can now calculate the variance of $\Delta N_A(t)$ - the difference between the actual $N_{A}(t)$ and the average number $K_A=Kx_A^*$ of organisms. We have:
\begin{widetext}
\bq
	\left<|\Delta N_A(t)|^2\right> = \frac{K_A^2}{2\pi} \int_{-\infty}^\infty \left<|\tilde{y}_{A}|^2(\omega)\right> d\omega = \frac{K_A^2}{2\pi} \int_{-\infty}^\infty \frac{(a_{BB}^2+\omega^2) \left<|\eta_A|^2(\omega)\right> 
		+ a_{AB}^2\left<|\eta_B|^2(\omega)\right>}{a_{AB}^2 a_{BA}^2+2 a_{AB} a_{BA} \omega ^2+a_{BB}^2 \omega ^2+\omega ^4} d\omega,
	\label{eq:DNA2}
\eq
\end{widetext}
in which we used Eq. (\ref{eq:xAspec}) and assumed no correlation between the noise $\eta_A$ and $\eta_B$ ($\left<|\eta_A\eta_B|^2\right>=0$).

In the case of Poisson replication, we put $\left<|\eta_A|^2\right>=\left<|\eta_B|^2\right>=(2b)(K/2)$ and evaluate the integral (\ref{eq:DNA2}) numerically. For our usual choice of the parameters $S_{0.5}$, we obtain $\left<|\Delta N_A(t)|^2\right>_{\rm theor}\approx 5.48\times 10^5$ which is very close to the numerical estimate from the simulation, $\left<|\Delta N_A(t)|^2\right>_{\rm sim}\approx 5.56\times 10^5$.

\begin{figure}
	\includegraphics[width=8cm]{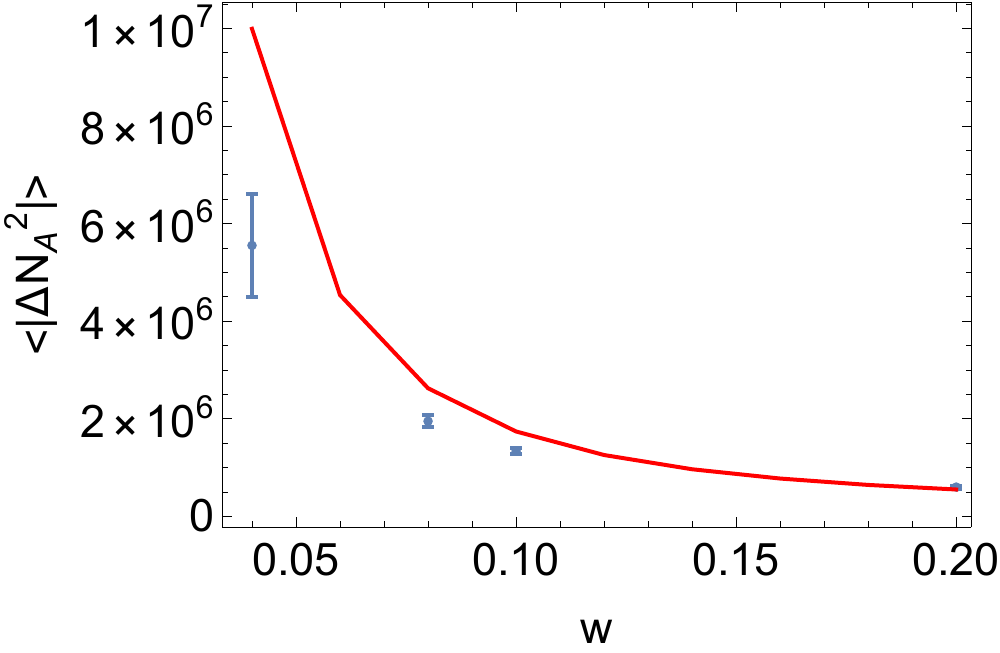}
	\caption{\label{fig:DNA2}Plot of $\left<|\Delta N_A(t)|^2\right>$ versus $w$, for the non-Poisson model with replication frequency $b=0.7$. Blue = simulation with $K=10^5$, and the remaining parameters as in $S_{0.5}$. Red = analytic expression \ref{eq:DNA2}.}
\end{figure}

In the case of quasi-synchronous replication, we insert Eqs. (\ref{eq:etafftfinal}) into Eq. (\ref{eq:DNA2}), and again integrate numerically over $\omega$. Figure \ref{fig:DNA2} shows the plot of $\left<|\Delta N_A(t)|^2\right>$ obtained in this way, compared to the simulation results. We notice that the analytic formula correctly reproduces the trend but the theoretically predicted values are generally larger than the ones from the simulation. However, the agreement is still quite good, given that our formula has been derived using many approximations.

Note that we used equation (\ref{eq:DNA2}), which is the same as the formula for the Poisson case \cite{mckane_predator-prey_2005}, but with coloured noise given by Eq. (\ref{eq:etafftfinal}) instead of white noise.
The fact that this approach works means that oscillations in the system with quasi-synchronous replication can be understood as being caused by resonant amplification of coloured, non-Poissonian noise.

To get some qualitative insight into the behaviour of Eq. (\ref{eq:DNA2}), we consider the case $\gamma\to 0$. For $T=1, b=\ln 2$, we can expand the formula under the integral in (\ref{eq:DNA2}) around $\omega=2\pi$, which  enables us to carry out the integral analytically. We obtain
\begin{widetext}
\bq
	\left<|\Delta N_A(t)|^2\right> \cong 
	\frac{KD_2\pi^2 (a_{BB}^2 + a_{AB}^2 + 4\pi^2) (\ln 2)^2}{\gamma (a_{AB}^2 a_{BA}^2 + 4 (2 a_{AB} a_{BA} + a_{BB}^2)\pi^2 + 
	16 \pi^4) (4 \pi^2 + (\ln 2)^2)} .
\eq
\end{widetext}
We see that, since $\gamma \sim w^2$, the variance of $\Delta N_A$ increases as $1/w^2$ as reproduction becomes more and more synchronous for $w\to 0$. This is similar to the effect of a long delay in reaction kinetics \cite{scott_long_2009}. The relationship $\left<|\Delta N_A(t)|^2\right> \propto K/\gamma$ can be interpreted as an effective reduction in the number of replicating entities; cells originating from a common ancestor replicate quasi-synchronously when their sub-population is much less than $1/\gamma$. The system thus consists of $K\gamma\ll K$ of such groups of cells, which increases demographic noise by a factor $1/\sqrt{\gamma}$, and the variance of $N$ by $1/\gamma$.

Let us now consider how small $w$ needs to be for the variance to start deviating from the Poisson case, i.e., how synchronous replication must be to make difference to random, asynchronous replication. Equation (\ref{eq:DNA2})
can be rewritten as follows:
\bq
  	\left<|\Delta N_A(t)|^2\right>
  = \frac{K_AD_2}{2\pi} \int_{-\infty}^\infty
  F(\omega) (f_0(\omega) + f_\gamma(\omega))d\omega, \label{eq:DNA3}
\eq
where
\bq
 F(\omega) = \frac{a_{BB}^2+a_{AB}^2 (K_B/K_A)+\omega^2}{(\omega^2-\Omega^2)^2+c}
\eq
is the resonance response function, with $c=-a_{AB} a_{BA} a_{BB}^2 - a_{BB}^4/4$, and $\Omega^2=-a_{AB}a_{BA}-a_{BB}^2/2=4\pi^2$ (squared resonant frequency for our parameters $S_{\rm 0.5}$). The function 
\bq
	f_0(\omega) = \frac{\omega^2}{\omega^2+(\ln 2)^2}
\eq
is the $\gamma$-independent contribution from stochastic replication, and the function
\bq
	f_\gamma(\omega)=\frac{ 
	2(\ln 2)^2 \omega^2}{(\gamma^2 + ( \omega-2\pi)^2) (4\pi^2 + 
	(\ln 2)^2)}
\eq
is the $\gamma$-dependent contribution.

$F(\omega)$ has full width at maximum height (FWHM) approximately equal to $\sqrt{c}$, whereas $f_\gamma(\omega)$ has FWHM equal to $\approx 2\gamma \approx 13.16 w^2 $. We expect that when the contribution from $f_\gamma(\omega)$ near the peak of $F(\omega)$ is larger than the contribution from $f_0(\omega)$, the variance of $N_A$ will be dominated by synchronous replication. For $\gamma<\sqrt{c}$, these contributions can be crudely estimated as follows:
\bq
	\int_{2\pi-\sqrt{c}/2}^{2\pi+\sqrt{c}/2}  f_0(\omega) d\omega \approx \sqrt{c},
\eq
and
\bq
	\int_{2\pi-\gamma}^{2\pi+\gamma}  f_\gamma(\omega) d\omega \approx	2/\gamma,
\eq
so that the contribution from synchronous replication becomes comparable to death-induced noise for $\gamma<2/\sqrt{c}$, or when $w< 0.55 c^{-1/4}$. As the expression is rather insensitive to the value of $c$, we can conclude that deviations from the Poisson, asynchronous replication should already be visible even for relatively large values of $w$.

\section{Conclusion}
We have revisited a stochastic two-species model of the predator-prey type \cite{mckane_predator-prey_2005}, which exhibits oscillations for a wide range of parameters of the model. We have modified the model so that both species replicate quasi-synchronously, with doubling times drawn from a narrow distribution. We have shown that coloured demographic noise generated by this process leads to much stronger oscillations than the Poisson process of replication assumed in earlier works. Coloured noise has been shown to affect population dynamics in single-species models \cite{spanio_impact_2017}; here, we not only derive its spectrum from the underlying microscopic dynamics, but also show how it affects more complex models.

Our result, while obtained for an abstract mathematical model, may be relevant for real biological populations, in particular for microorganism which often replicate in quasi-discrete generations. We expect to see the same behaviour in other models that exhibit quasi-cycles \cite{prigogine_symmetry_1968,xia_transient_2005,lemesle_simple_2008, garai_stochastic_2012, dobrinevski_extinction_2012, gavagnin_synchronized_2021}.

The phenomenon of coloured noise amplification may be further augmented in situations in which oscillations in the population abundance become synchronized with reproductive cycles. We leave this interesting problem for future studies.

\section*{Acknowledgments}
B.W. acknowledges funding under Dioscuri, a programme initiated by the Max Planck Society, jointly managed with the National Science Centre in Poland, and mutually funded by Polish Ministry of Science and Higher Education and German Federal Ministry of Education and Research (UMO-2019/02/H/NZ6/00003).

\bibliographystyle{apsrev-nourl}
\bibliography{literature}

\end{document}